\documentclass{pasj01}
\usepackage{mathrsfs} 
\usepackage{graphicx} 
\usepackage{mediabb}
\usepackage{color}  
  
\title{FOREST Unbiased Galactic Plane Imaging Survey with the Nobeyama 45-m Telescope (FUGIN) IV:
 Galactic Shock Wave and Molecular Bow Shock in the 4-kpc Arm of the Galaxy}
 
\author{Y. \textsc{Sofue}\altaffilmark{1}, 
M. \textsc{Kohno}\altaffilmark{2 }, 
K. \textsc{Torii}\altaffilmark{3},
T. \textsc{Umemoto}\altaffilmark{3}, 
N. \textsc{Kuno}\altaffilmark{5,6}, 
K. \textsc{Tachihara}\altaffilmark{2 }, 
T. \textsc{Minamidani}\altaffilmark{3,4} , 
S. \textsc{Fujita}\altaffilmark{2,3,5} , 
M. \textsc{Matsuo}\altaffilmark{3,7} , 
A. \textsc{Nishimura}\altaffilmark{2} , 
Y. \textsc{Tsuda}\altaffilmark{8},
M. \textsc{Seta}\altaffilmark{9}
}

\altaffiltext{1}{Inst. Astronomy, The Univ. of Tokyo, Mitaka, Tokyo 181-0015}
\altaffiltext{2}{Dept. Astrophys., Nagoya Univ.,  Chikusa-ku, Nagoya 464-8602}
\altaffiltext{3}{Nobeyama Radio Obs., NAOJ, Nobeyama, Nagano 384-1305}
\altaffiltext{4}{Dept. Astron. Sci., Sch. of Phys. Sci., Mitaka, Tokyo 181-8588}
\altaffiltext{5}{Dept. Phys., Grad. Sch. of Pure and Appl. Sci., Univ. of Tsukuba,  Ibaraki 305-8577}
\altaffiltext{6}{Tomonaga Cen. History of the Universe, Univ. of Tsukuba, Ibaraki 305-8571}
\altaffiltext{7}{Grad. Sch. of Sci. and Engi., Kagoshima Univ., Kagoshima 890-0065}
\altaffiltext{8}{Dept. Physics, Grad. Sch. of Sci. and Tech., Meisei Univ., Hino, Tokyo 191-0042}
\altaffiltext{9}{Dept. Physics, Kwansei Gakuin Univ., Gakuen 2-1, Hyogo 669-1337}

\email{sofue@ioa.s.u-tokyo.ac.jp}


\begin{document} 
\date{ } 
\maketitle   

\def\vlsr{v_{\rm LSR}} \def\v{v_{\rm LSR}} \def\Msun{M_\odot} \def\deg{^\circ} \def\r{\bibitem[]{}}     \def\/{\over}\def\kms{km s$^{-1}$}  \def\Vsun{V_0}  \def\Vrot{V_{\rm rot}}   \def\Tc{T_{\rm C}} \def\Tb{T_{\rm B}} \def\sin{{\rm sin}\ } \def\cos{{\rm cos}\ } \def\Hcc{ H cm$^{-3}$ }  \def\co{$^{12}$CO$(J=1-0)$ }\def\co{$^{12}$CO$(J=1-0)$ } \def\htwo{H$_2$} \def\sub{\subsection}\def\be{\begin{equation}} \def\ee{\end{equation}}\def\Kkms{K \kms} \def\Htwosqcm{H${_2}$ cm$^{-2}$}\def\Icut{I_{\rm c}} \def\Ico{I_{\rm CO}} 
\def\Icri{I_{\rm c}} \def\Ic{I_{\rm c}}
\def\Xco{X_{\rm CO}} \def\X{X_{\rm mass}} 
\def\XHI{X_{\rm HI}} \def\TbHI{T_{\rm b: HI}} \def\IHI{I_{\rm HI}}\def\Te{T_{\rm e}}\def\ne{n_{\rm e}}\def\ue{u_{\rm e}} \def\um{u_{\rm mol}}\def\ergcc{erg cm$^{-3}$}\def\nHH{n_{\rm H_2}}\def\mH{m_{\rm H}}
\def\sigv{\sigma_v} \def\x{\times}\def\Hcc{H cm$^{-3}$} \def\fmol{f_{\rm mol}}
\def\tauco{\tau_{\rm CO}}\def\tauhi{\tau_{\rm HI}}\def\Ts{T_{\rm S}}\def\Tco{T_{\rm B: CO}}\def\THI{T_{\rm B: HI}}\def\nH2{n_{\rm H2}}\def\nHI{n_{\rm HI}}\def\Halpha{ H$\alpha$ }\def\nuv{N_{\rm UV}}\def\ne{n_{\rm e}}\def\ni{n_{\rm i}}\def\nh{n_{\rm H}}\def\ar{\alpha_{\rm r}}\def\cs{c_{\rm s}}\def\Luv{L_{\rm uv}}\def\RS{R_{\rm S}}\def\rhii{R_{\rm HII}}\def\Lsun{L_\odot}\def\subsub{\subsubsection}
\def\fmol{f_{\rm mol}} \def\noi{\noindent}
\def\Htwo{H$_2$ } \def\sigv{\sigma_v }
\def\bf{ }

\begin{abstract}
The FUGIN CO survey revealed the 3D structure of a galactic shock wave in the tangential direction of the 4-kpc molecular arm. 
The shock front is located at G30.5+00.0+95 \kms on the up-stream (lower longitude) side of the star-forming complex W43 (G30.8-0.03), and composes a molecular bow shock (MBS) concave to W43, exhibiting an arc-shaped molecular ridge perpendicular to the galactic plane with width $\sim 0\deg.1 \ (10\ {\rm pc})$ and vertical length $\sim 1\deg \ (100\ {\rm pc})$. 
The MBS is coincident with the radio continuum bow of thermal origin, indicating association of ionized gas and similarity to a cometary bright-rimmed cloud. 
The up-stream edge of the bow is sharp with a growth width of $\sim 0.5$ pc indicative of shock front property. 
The velocity width is $\sim 10$ \kms, and the center velocity decreases by $\sim 15$ \kms from bottom to top of the bow. The total mass of molecular gas in MBS is estimated to be $\sim 1.2\times 10^6 \Msun$ and ionized gas $\sim 2\times 10^4 \Msun$. 
The vertical disk thickness increases step like at the MBS by $\sim 2$ times from lower to upper longitude, which indicates hydraulic-jump in the gaseous disk. We argue that the MBS was formed by the galactic shock compression of an accelerated flow in the spiral-arm potential encountering the W43 molecular complex.
A bow-shock theory can well reproduce the bow morphology. We argue that molecular bows are common in galactic shock waves not only in the Galaxy but also in galaxies, where MBS are associated with giant cometary HII regions. 
We also analyzed the HI data in the same region to obtain a map of HI optical depth and molecular fraction. We found a firm evidence of HI-to-\htwo transition in the galactic shock as revealed by a sharp molecular front at the MBS front.
\end{abstract}   

\section{Introduction}
  
  A galactic bow shock was observed in the tangential direction at G30.5+00 of the 4-kpc molecular arm (Scutum arm) in radio continuum emission at 10 GHz using the Nobeyama 45-m telescope (Sofue 1985), which revealed a cross section of the gaseous spiral arm. The radio bow shock showed a concave arc with respect to the star forming (SF) region W43 at G30.8-0.03. (Hereafter, G$l\pm b$ represents the position or name of the object centered at $l$ and $b$ in degrees.)
  
   From the concave structure, it was interpreted as due to a bow shock produced by a supersonic flow of interstellar gas encountering the massive molecular complex around W43. A molecular arc in the \co line at $\vlsr=90 -100$ \kms was also found to be associated with the radio bow shock using an earlier low resolution CO observations. The molecular clouds and bow shock G30.5 are considered to be located near the tangent point of the 4-kpc arm (Scutum arm) outside the supposed Galactic bar.
   
Astrophysical bow shock is a classical subject, and is commonly observed associated with objects interacting with supersonic flows. There have been a number of observations and models in various-scaled objects, from geomagnetosphere in the solar wind (Baranov et al. 1971), interstellar globules against ISM flow and UV field (Dyson et al. 1975), shocks in HII regions (van Buren et al. 1990; Povich et al. 2008), stellar winds against ISM flow (Wilkin 1996; Ueta et al. 2008; Tarada 2012), stellar outflows and environment (Arce and Goodman 2002), and cosmic jets against ambient medium (Ogura et al. 1995; Reipurth et al. 2002; Sakemi et al. 2018 ).

In this paper we revisit the galactic bow shock G30.5+00 in the 4-kpc arm (Scutum arm), and study its detail using the most recent CO-line observations. We model the bow-shock as due to supersonic flow of gas in the galactic-shock wave encountering the star forming complex W43. We argue that the galactic-arm scale bow shock is a common phenomenon in spiral arms of galaxies observed as giant cometary HII regions. We discuss these topics based on the galactic shock wave theory (Fujimoto 1969;  Roberts et al. 1972;  Tosa 1991; Mishurov 2006).

As to the SF activity in the W43 region, a number of papers have been published based on observations from radio to infrared emissions (Liszt 1995; Liszt et al. 1993; Subrahmanyan \& Goss 1996; Fux 1999; Bally et al. 2010; Nguyen-Lu'o'ng et al. 2011; Carlhoff et al. 2013; Beuther et al. 2012; Motte et al. 2014; Bihr et al. 2015; Saral et al.(2017)). 
The FUGIN CO survey has provided higher resolution and more sensitive maps in the CO lines in a wide field around W43, and revealed detailed structures and kinematics of the molecular complex associated with W43. Results of full analyses of the FUGIN  data of the W43 region are presented in Kohno et al. (2018b) along with a detailed study of SF activity.

For analysis, we use the \co line data from the  FUGIN CO-line survey (Umemoto et al. 2017;  Kohno et al. 2018).  
 The data cubes covered the galactic plane region around W43 from $l=29\deg$ to $32\deg$ and $b=-1\deg$ to $+1\deg$ from FUGIN that had spatial and velocity resolutions of $20''\times 20''\times 0.65$ \kms.
{\bf We use only the \co data, because extended molecular feature is more clearly observed than other CO lines and the mass estimation will be made using the CO-to-\Htwo conversion factor available only for \co. }

VLBI parallax measurements of maser sources (Zhang et al. 2014; Sato et al. 2015) in G31.5 and G29.96 SF regions (not W43 Main at G30.8) indicated a mean distance of $r\sim 5.5 $ kpc. This distance agrees with a near-side kinematical distance corresponding to $\v \sim 95$ \kms at $l\sim 30\deg.8$ for the galactic rotation velocity (Sofue 2013) with $V\simeq 225$ \kms at $l\sim 30\deg$ and solar constant of $V_0=238$ \kms (Honma et al. 2013). We adopt this distance for W43 and associated molecular clouds at $\vlsr \sim 95$ \kms, as well as the galactic bow shock G30.5+00. We assume a galacto-centric distance of the Sun of $R_0=8.0$ kpc, which locates W43 at a galacto-centric distance of 4.3 kpc at an azimuthal angle of $40\deg.6$.

\section{The Maps}

\subsection{Continuum, CO and HI maps}

\begin{figure*} 
\begin{center}     
\includegraphics[width=13cm]{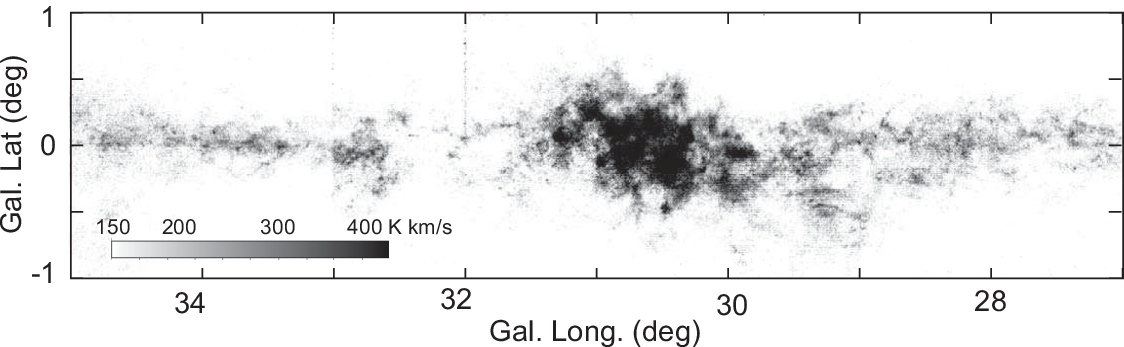}     
\end{center}
\caption{\co intensity map integrated from $\vlsr=-20$ to $+160$ in the tangential direction of the 4-kpc arm around W43 (G30.8). }
\label{wide}
	\end{figure*}

{\bf
Figure \ref{wide} shows an integrated intensity map of the \co line emission of the galactic plane at $l=27\deg$ to $35\deg$ integrated at $\vlsr=-20$ to 160 \kms. The tangential direction of the 4-kpc arm (Scutum arm) including the star forming complex W43 is visible as a large clump around G31, showing the cross section of the arm. The gray scaling starts at $\Ico=150$ \Kkms, so that the map represents relatively high column regions near the tangent points of the galactic rotation.
}

\begin{figure} 
\begin{center}     
(a)\includegraphics[width=7cm]{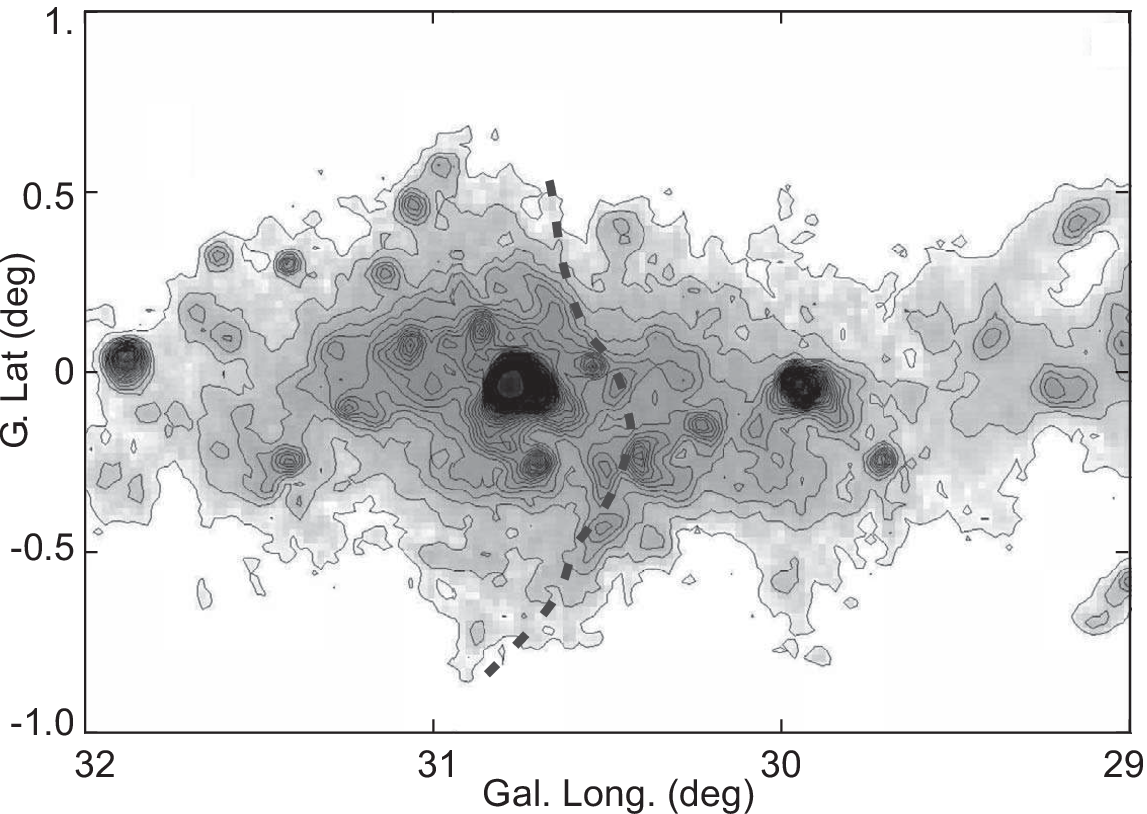}      
(b)\includegraphics[width=7cm]{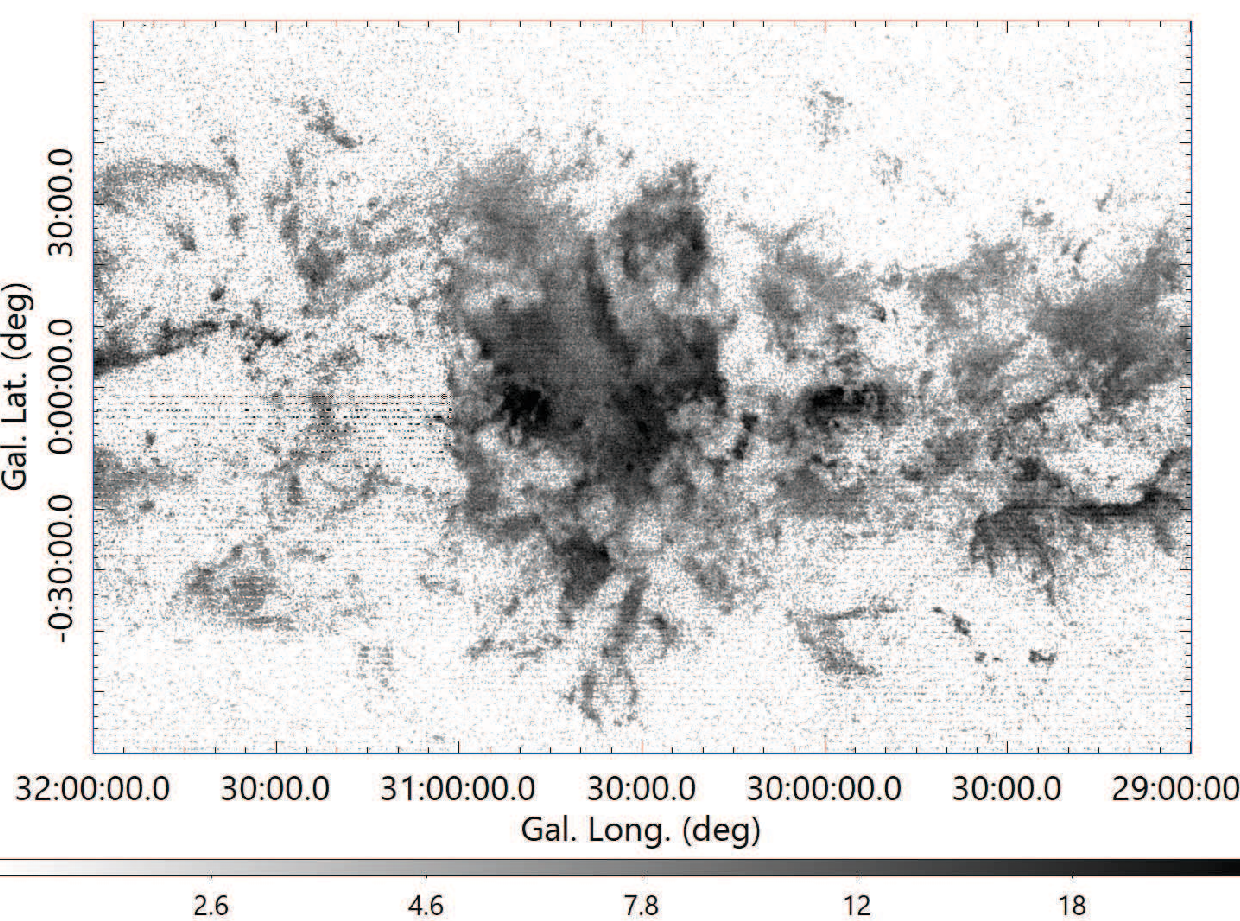} 
(c)\includegraphics[width=7cm]{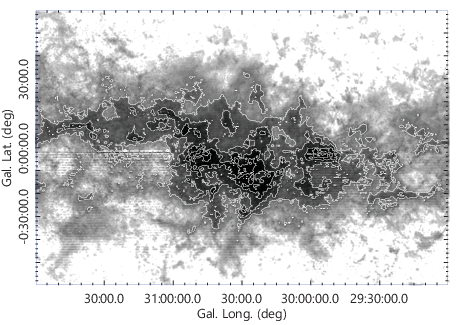}  
(d)\includegraphics[width=7cm]{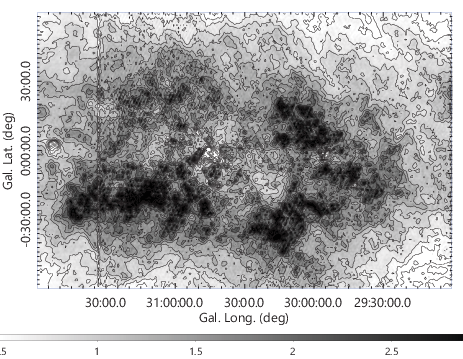}  
\end{center}
\caption{  
{\bf (a) 10GHz continuum map (Sofue 1985; Handa eta l. 1987), showing HII regions W43 Main (G35.8) and West (G29.96), and a radio bow shock at G30.5 concave to W43 Main. The dashed line traces a molecular ridge in (b).
(b) A channel map of \co  line  at  $v=94.465$ \kms. 
(c){\bf  \co integrated intensity map from  $\v=80$ to 120 \kms. Contours are at $\Ic=100$ \Kkms (approximately surrounding GMA (see text)); 200 (GMC); 300 (DMC) and 400 \Kkms (cores).  Horizontal stripes at $l>31\deg$ and $b<0\deg$ are instrumental artifacts. } 
(c) HI optical depth, $\tauhi$, calculated for averaged $\Tb$ at $\v=85$ to 105 \kms from THOR HI survey by Bihr et al. (2015).  
Values toward strong continuum sources brighter than $\Ts=130$ K (W43, G29.96 etc.) are not valid. The vertical stripe at $l=30\deg.1$ is an artifact.
}
}
\label{maps} 
	\end{figure}

In figure \ref{maps} we compare the CO map around W43 with those in radio continuum and HI line.
The radio continuum map at 10 GHz observed with the Nobeyama 45-m telescope (Sofue 1985; Handa et al. 1987) in figure \ref{maps}(a) reveals the radio bow shock as a vertical arc extending from G30.5+00 toward both latitudes up to $b \sim \pm 0\deg.7-0\deg.8$. The arc is concave to the strong radio source W43 Main at G30.8-0.03. The dashed line traces the sharpest CO ridge recognized in the CO channel map at $\v=95$ \kms.

 Figures \ref{maps}(b) and (c) show a CO channel map at $\vlsr=95$ \kms and intensity integrated from $\v=85$ to 105 \kms with cutting flux at 2 K.  
 Three dense molecular regions are evident at G30.0 near the compact HII region G29.96 (W43 West), G30.5 (molecular bow shock) and G30.8 (W43 Main). Besides these clouds, vertically extended arcs are visible at G30.0, G30.5 and G31.6, particularly prominent at G30.5. We focus on this long arc at G30.5, and call it the molecular bow shock (MBS) G30.5. 
The MBS  positionally coincides with the radio continuum bow, extending toward latitudes to $b\sim \pm 0\deg.8$, and is concave to W43. 

 Figure \ref{maps}(d) shows optical depth map of the 21-cm HI line for averaged brightness temperature between 85 and 105 \kms for an assumed spin temperature of 130 K, as obtained using THOR HI survey by Bihr et al. (2015). There appears a significantly deficient region of HI around G30.5 in positional coincidence with the molecular bow. In the channel maps of HI emission, this deficiency is visible in the velocity range from 90 to 100 \kms close to that of MBS. The velocity-selected deficiency confirms that the HI deficiency is not due to absorption against continuum background, but HI gas is really less abundant in this velocity range. This feature will be discussed in relation to HI to \htwo transition in section \ref{discussion}.

\subsection{Molecular gas components} 
        
{\bf We categorize the molecular structures in the analyzed region into four components using the integrated intensity map in analogy to usual ISM terms: (i)  giant molecular association (GMA),  
(ii) giant molecular clouds (GMC), (iii) dense molecular clouds (DMC), and (iv) molecular cores. We define these four components as clouds in regions enclosed by contours at threshold intensities at $\Ic=100$ \Kkms, 200, 300, and 400 \Kkms, respectively, in the \co intensity map integrated between $\v=80$ and 120 \kms shown in figure \ref{maps}(c).  
} 

The mass of each component is measured by surface-integration of the intensity enclosed by the threshold contours at $\Ico=\Ic$ as
\be
M=\X \int \Ico(\ge \Ic) dx dy,
\label{eqmass}
\ee
where $\X$ is the CO to \htwo conversion factor, $dx$ and $dy$ are linear extents in the $l$ and $b$ direction, and we assume a distance of 5.5 kpc.
{\bf  
We adopt a conversion factor for the \co line, 
$X_{\rm CO}=2.0\times 10^{20} {\rm  H_2 cm^{-2} [K\ km\ s^{-1}]^{-1}}=3.20\Msun {\rm pc^{-2} [K\ km\ s^{-1}]^{-1}}$,
which yileds $\X=4.3 \Msun {\rm pc^{-2}  [K\ km\ s^{-1}]^{-1}}$ including heavy elements (Bolatto et al. 2013).
}
We list the calculated masses in table \ref{tabmass}.

	\begin{table*}  
\caption{Sizes$^\dagger$, masses$^*$, energies and Jeans times of molecular clouds$^\ddagger$  calculated for $\sigma_v=10$ \kms.
}
\begin{tabular}{lllllllllll}  
\hline 
\hline   
Compo.$^\dagger$ & $\Ic$ & $\langle {\Ic} \rangle$ & $D$ & $M$ &$\sigv$& log $E_{\rm g}$ & log $E_{\rm k}$ & $E_{\rm g}/E_{\rm k}$ & $n_{\rm H_2}$ & $t$  \\
  &\small{(\Kkms)} &\small{(\Kkms)}& (pc)& ($\Msun$) &(\kms)& (erg) & (erg)& & \small{(H$_{2}$ cm$^{-3}$)} & (My)  \\
\hline
 GMA &  100.0&  168.3&  120.7& 0.13E+08&    7&  53.29&  51.81&   29.9& 0.11E+03&   5.53\\
HI$^\#$ & -- & -- & -- & 0.17E+07  &  -- & -- & -- & -- &0.75E+02H & --  \\
 GMC &  200.0&  259.2&   58.0& 0.47E+07&    8.5&  52.71&  51.52&   15.3& 0.35E+03&   3.09\\
 DMC &  300.0&  345.0&   25.6& 0.12E+07&   11&  51.89&  51.19&    5.1& 0.10E+04&   1.78\\
 Core &  400.0&  458.5&    8.2& 0.16E+06&   11&  50.65&  50.32&    2.2& 0.44E+04&   0.87\\ 
\hline
MBS & & & Vol. (pc$^3$)\\
\hline 
\ \ Mol. &-- & 150 & \small{$100^2\times 10$} &0.12E+07 &--& --&--&--& 0.13 E+03&--\\
\ \  HII &-- & -- & -- &0.20  E+05 &--&--& 49.5 (th.) &--&  8 H &--\\
\hline
\end{tabular} \\
$^*$ $\X=4.3\Msun {\rm pc^{-2} [K\ km\ s^{-1}]^{-1}}$ (including heavy elements: 1.36 times the \Htwo mass for $X_{\rm CO}=2.0\times 10^{20} {\rm  H_2 cm^{-2} [K\ km\ s^{-1}]^{-1}}$).\\
$^\ddagger$ GMA: giant molecular association; GMC: giant molecular cloud; DMC: dense molecular cloud; MBS: molecular bow shock.\\
$^\dagger$ Mean values for each component can be estiamted by $D'\sim D/\sqrt{N}$, $M'\sim M/N$, $E'_{\rm g}\sim E_{\rm g}/\sqrt{N}$,  $\rho'\sim \sqrt{N} \rho,$ and $t'\sim N^{-1/4}t$ (not listed here), where  $N\sim 1$ for GMA, and $\sim 3$ for GMC and DMC/cores. \\
$^\#$ HI mass $\sim A\times \Sigma_{\rm HI}$, where $\Sigma_{\rm HI}\sim 150 \Msun {\rm pc}^{-2}\simeq 1.9\times 10^{22}\ {\rm H\ cm}^{-2}$ (Bihr et al. 2015).
\label{tabmass}  
	\end{table*}    
        
It may be stressed that the here estimated mass of individual GMC, $M\sim 1.5\times 10^6\Msun$, (see below) is comparable to that of usual GMC. 
It should be also noted that the mass of the cores $\sim 1.6\times 10^5\Msun$ in diameter of $\sim 8$ pc is consistent with that obtained for the W43 Main+West clouds of $\sim 2\times 10^5\Msun$ in $\sim 20\times 10$ pc$^2$ from dust emission (Lin et al. 2016).

As for sizes, we define the representative diameter (size) and radius, $D=2a$, by $A=\pi a^2=\pi D^2/4$,
 where $A$ is the area enclosed by the threshold contours in figure \ref{maps}(c). 
 In table \ref{tabmass} we list the measured values, where $\Ic$ is the threshold intensity and $\langle \Ic \rangle$ is the averaged intensity in area $A$. 

{\bf
We also estimate the gravitational energy by $E_{\rm g}=GM^2/a$, assuming that the line-of-sight extent is equal to the diameter, where $G$ is the gravitational constant. We compare it with the kinetic energy $E_{\rm g}\sim M \sigv^2/2$ in order to confirm that $E_{\rm g}/E_{\rm k}>1$ so that the clouds are gravitationally bound. The velocity dispersion was estimated by $\sigv \sim \sqrt{2}w$, where $w$ is the velocity width in the moment-2 map. Using figure \ref{mom12}(b) we estimated the width as $w \sim 5$ \kms for GMA, $\sim 6$ \kms for GMC, and $\sim 8$ for DMC and cores, and hence, $\sigv \sim 7,\ 8.4,$ and 11 \kms, respectively.  
}
The mass is further used to calculate the density by $\rho=M/(4\pi a^3/3)$ and the Jeans time $t=1/\sqrt{G\rho}$ (time scale of gravitational contraction).In table \ref{tabmass} we list the calculated values for individual components.

If there exist $N$ clouds in the analyzed field, individual diameter of each component is given by $D'\sim D/\sqrt{N}$. Accordingly, the mass of a single typical cloud is estimated by $M' \sim M/N$. As the total gravitational energy is overestimated by a factor of $\sim \sqrt{N}$, individual energy is reduced to $E'_{\rm g}$ and related to $E_{\rm g}$ by
$E'_{\rm g}\sim E_{\rm g}/\sqrt{N}.$
Similarly, individual density $\rho'$ and Jeans time $t'$ are approximated by the calculated density   and   time by
$\rho'\sim \sqrt{N} \rho,$
and
$t'\sim N^{-1/4}t.$ 
By eye estimates in figure \ref{maps}(c), we counted the numbers of clouds to be $N\sim 1$ for the GMA, $\sim 3$ for GMC, DMC and cores.

\subsection{Kinematical properties}

\begin{figure} 
\begin{center}      
(a)\includegraphics[width=7cm]{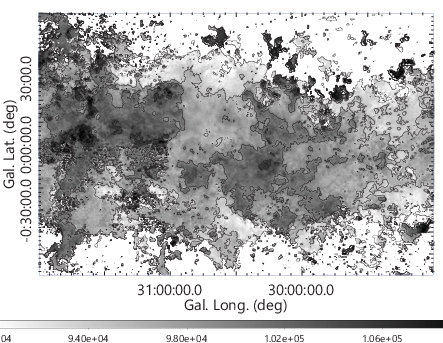} 
(b)\includegraphics[width=7cm]{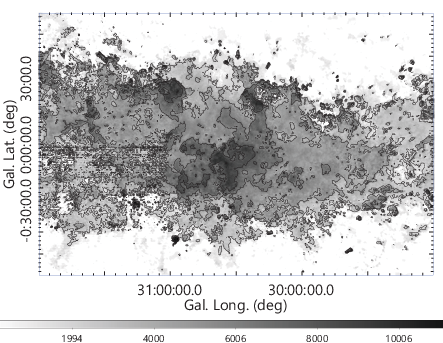} 
\end{center}
\caption{(a) Moment 1 (velocity field) map with contours every 5 \kms. Note the contours running almost vertically, showing systematic variation with longitude, but inversely against the galactic rotation near G30.5. (b) Moment 2 (velocity width) with contours every 2 \kms.}
\label{mom12}
	\end{figure}

 Figures \ref{mom12}(a) and (b) show moment 1 and 2 maps, showing the intensity-weight velocity field and velocity width distributions, respectively. The velocity field shows systematic variation with longitude. It shows a rapid decrease in velocity at G30.5 with gradient much steeper than the decrease due to the galactic rotation, indicating sudden deceleration of gas near G30.5.
 
Although it shows the general variation of velocities, moment 1 map is contaminated by widely extended components at velocities between 80 to 120 \kms. In order to examine the velocity structure more specifically for the MBS, we made a color coded intensity map around G30.5 using channel maps closer to the MBS's center velocity at 91, 95 and 99 \kms in blue, green and red, respectively.

\begin{figure} 
\begin{center}    
(a)\includegraphics[width=7cm]{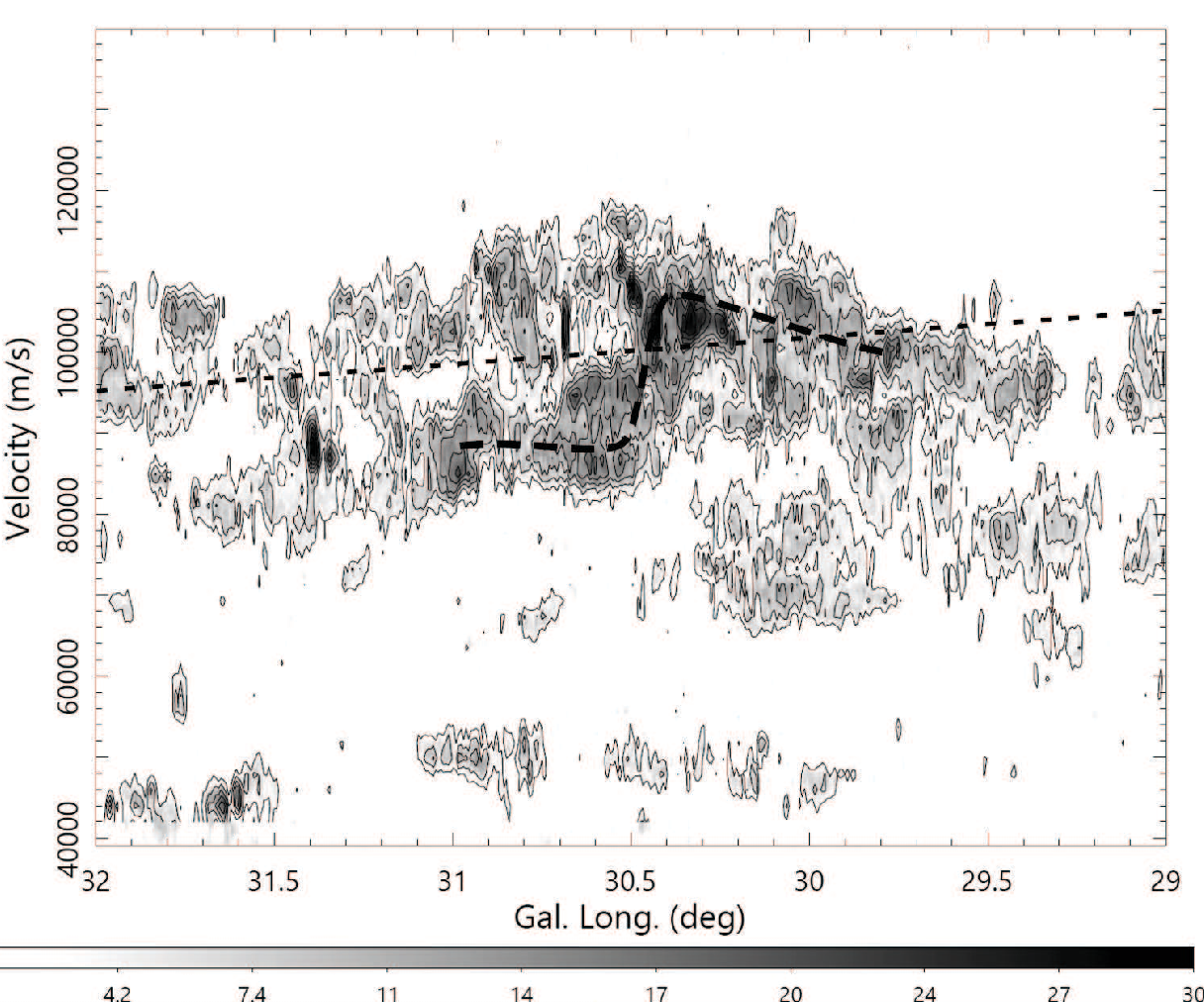}  
(b)\includegraphics[width=7cm]{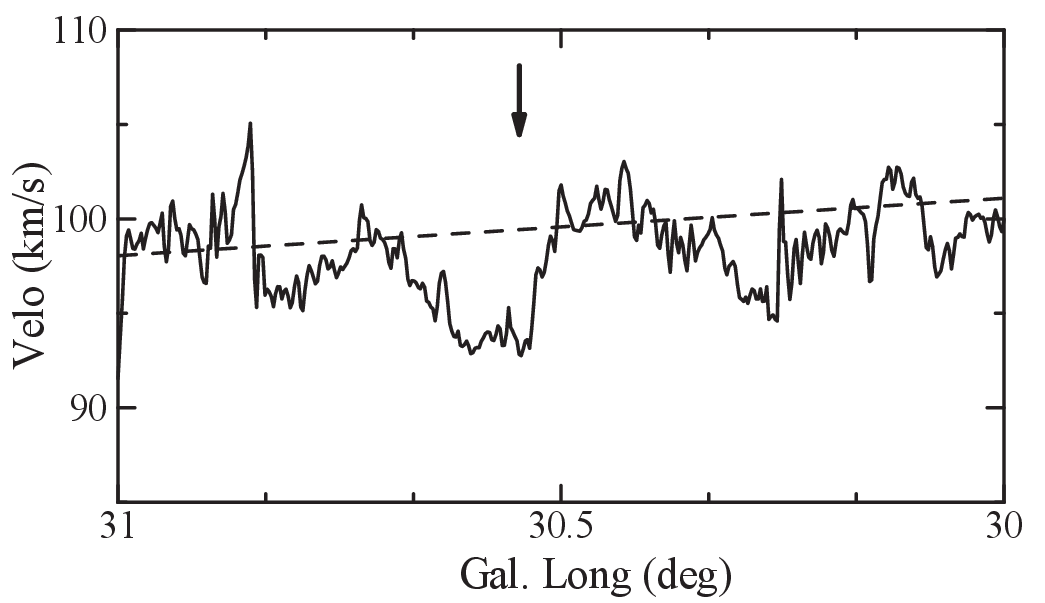}   
\end{center}
\caption{(a) LV diagram at $b=-0\deg.25$. Contours are drawn every 3 K starting at 3 K. Thick and thin dashed lines indicate the trace of the velocity jump from up- to down-stream side of the bow, and terminal velocity expected from the galactic rotation, respectively. (b) Variation of intensity weighted mean $\v$ along $b=-0\deg.25$, showing sudden deceleration of radial velocity at the MBS (arrow). The dashed line represents the smoothed galactic rotation curve. }
\label{lv_jump}
	\end{figure}

The G30.8+00 molecular complex associated with the HII region W43 is kinematically located near the tangent point along the 4-kpc molecular arm (Scutum arm) at radial velocity slightly slower than the terminal velocity.  
Figure \ref{lv_jump}(a) shows an LV diagram at $b=-0\deg.25$, where the velocity of molecular gas around the MBS changes drastically from the lower to upper longitudes as indicated by the thick dashed line. Panel (b) shows the variation of intensity weighted mean $\v$ (moment 1 as in figure \ref{mom12}) at the same latitude from $l=30\deg$ to $31\deg$, showing a sudden decrease in the radial velocity at the MBS (arrow).
The velocity variation is much steeper than expected from the terminal velocity due to the galactic rotation as shown by the thin dashed line.
 
\begin{figure} 
\begin{center}    
\includegraphics[width=6cm]{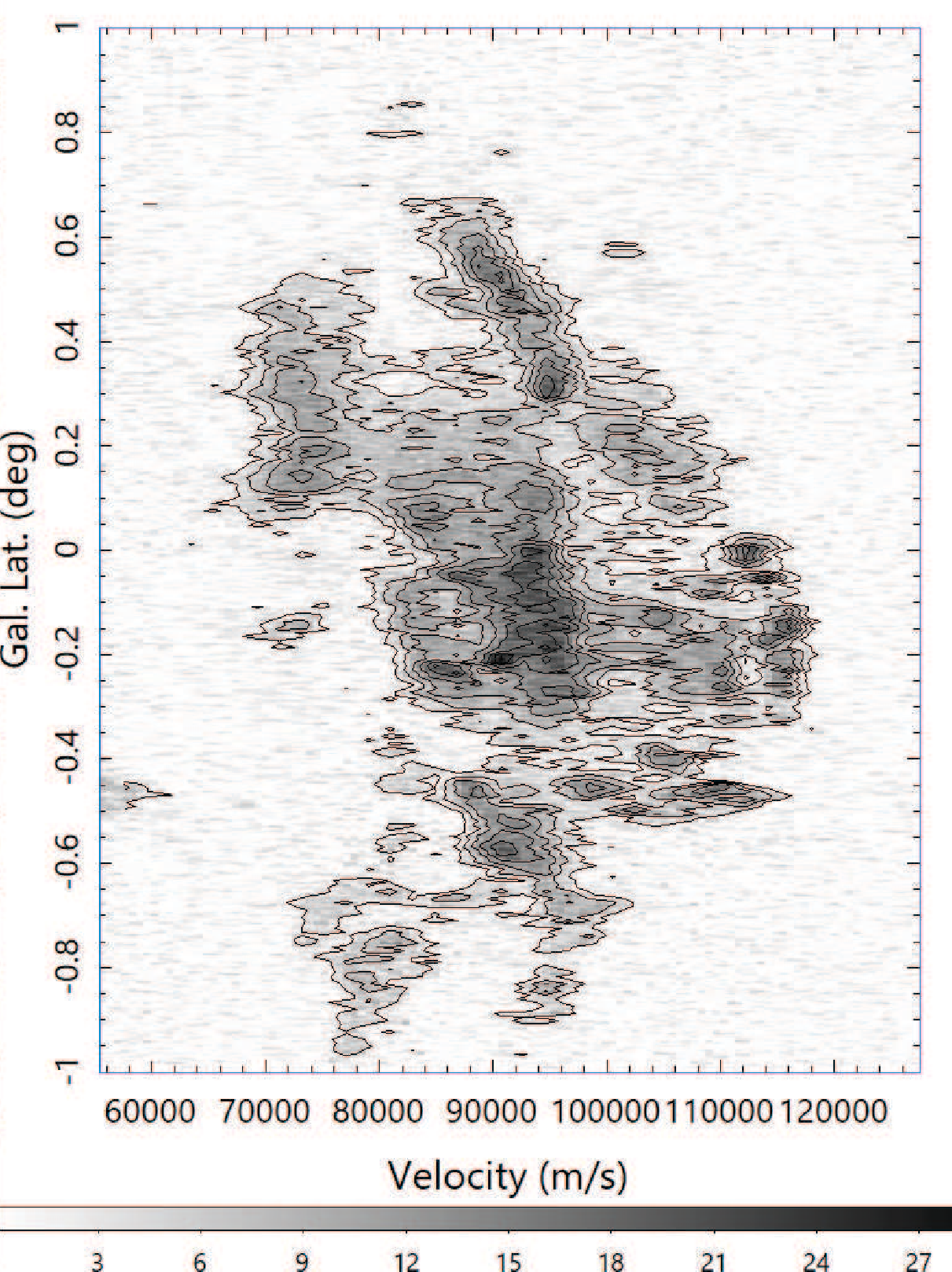}   \\
\includegraphics[width=4cm]{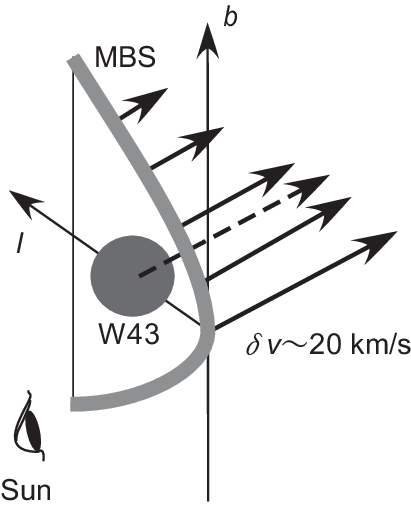} 
\end{center}
\caption{(Top) Velocity-latitude diagram at G30.513 . Contours are drawn every 3 K starting at 3 K. 
{\bf (Bottom) Schematic view of relative line-of-sight motion of the MBS's tangent ridge.}}
\label{vb_enlarge}
\label{illust_vb}
	\end{figure}

In order to separate the kinematical behaviors from the surrounding clouds, we made velocity-latitude diagrams at different longitudes around the MBS. The MBS gas is extended in the latitude direction for more than $\pm 0\deg.9$ (90 pc), or further beyond the observed edges. Impressive in the VB diagrams is the bow-like edge of the $v - b$ profile concave to W43 molecular complex in the highest brightness region. We show a $v - b$ diagram at $l=37\deg.5$ in  figure \ref{vb_enlarge}. 

Combining these kinematical behaviors in the $v-b$ diagrams with the geometrical and morphological consideration in the previous section, we may summarize the three dimensional kinematical structure of the molecular bow shock as illustrated in figure \ref{illust_vb}.

\section{3D Molecular Structure}

\subsection{Molecular bow}

In the \co intensity map, the molecular bow shock (MBS) is recognized as the vertical molecular arc extending from G30.5+00 toward both sides of the Galactic plane. The bow is concave with respect to W43 Main, and is visible in a narrow range of velocity from $\v \sim 90$ to 100 \kms, having full velocity width of $\sim 10$ \kms.

The extent of the MBS is $\Delta l\times \Delta b \sim 0.1\deg \times 1\deg$, which corresponds to $L\times W \sim 10 {\rm pc}\times 100 {\rm pc}$ for an assumed distance of 5.5 kpc, where $L$ is the vertical length and $W$ is the width. From the figures we estimate the mean intensity along the MBS to be $\Ico \sim 150$ \Kkms, which yields the total mass of molecular gas in the bow shock to be
$M_{\rm MBS}\sim 8.7\times 10^5 \Msun$. This shares about $\sim 0.19$ of the total GMC mass around W43 of $\sim 4.7\times 10^6 \Msun$ in table \ref{tabmass}. 

\begin{figure}  
\begin{center}     
 \includegraphics[width=8cm]{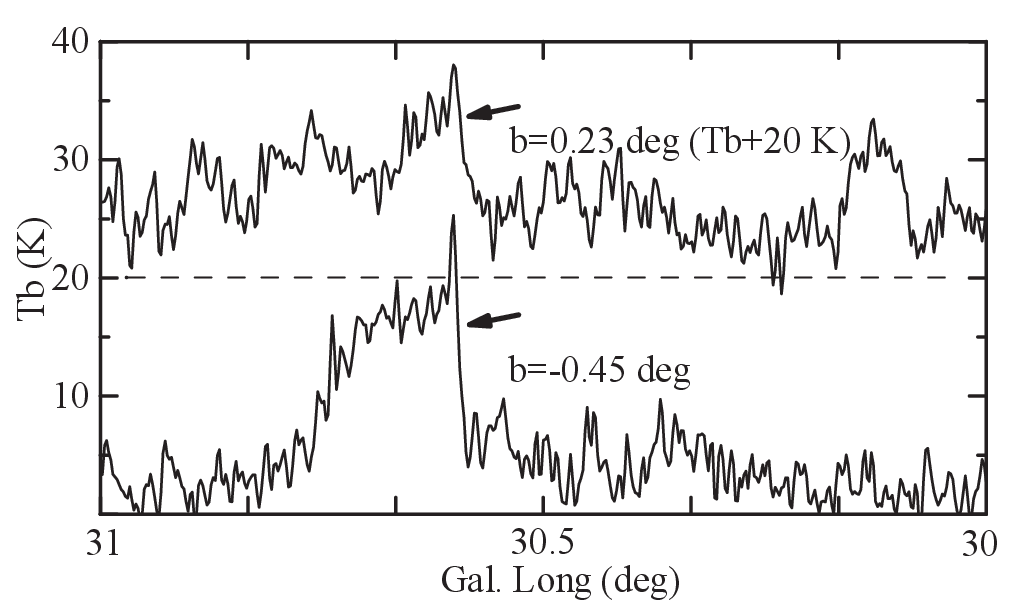}  
\end{center}
\caption{Horizontal cuts of the CO-line brightness at $v=95$ \kms at $b=-0\deg.45$ and $+0\deg.23$. Note the sharp intensity rise at the western edge. Triangles point longitudes of G29.96, G30.5 and W43 Main.}
\label{cut}
	\end{figure}

Figure \ref{cut} shows the horizontal variations of the brightness temperature at $v=94.675$ \kms from G30 to G31 at two different latitudes, $b=-0\deg.43$, $0\deg$, and $+0\deg.23$. The cross sections of the MBS indicated by the arrows exhibit a sharp edge with sudden increase of the intensity from right to left. The edge's growth width is comparable to the observing resolution of $20''$, or $\sim 0.5$ pc (figure \ref{cut}). 

Such cross sections are typical for shock wave compression. The bow-shaped sharp edge along the MBS may be the first clear evidence for a galactic shock wave in the Galaxy, which is caused by a supersonic flow from the up-stream side of the galactic rotation toward down-stream side against the 4-kpc arm.  

\subsection{Bow shock theory }

We point out that the radio continuum bow along MBS is similar to sub-pc scale bright-rimmed clouds (BRC) associated with thermal radio emission, where the ionization occurs due to UV photons from nearby forming stars (Urquhart et al. 2006; Sugitani et al. 1991; Thompson et al. 2004). Although the MBS is caused by an encounter with the supersonic flow in galactic rotation, the similarity suggests that the MBS would be a giant bright-rimmed molecular cloud (GBRC) of spiral-arm scale.  

Such an inside bright rim may be indeed the case in MBS G30.5, illuminated by UV photons from W43. In fact, the 10 GHz radio continuum emission in MBS G30.5 (figure \ref{maps}) is located slightly inside the molecular bow, and is possible that the emission is a mixture of ionized gas heated by the shock wave from outside and that excited by W43 from inside.
 
According to Wilkin's (1996) analytical model for stellar bow shock , the bow front shape is expressed as follows.
\be
Q(\phi)=\RS\ {\rm cosec}\ \phi \sqrt{3(1-\phi\ {\rm cot}\ \phi}),
\ee
where $Q$ and $\phi$ are the radius and elevation angle of the front from the galactic plane, $\RS$ is the stand-off distance defined as the distance of the front on the galactic plane from the central body responsible for triggering the shock wave. We here consider that $\RS$ is approximated by the distance of MBS at G30.5+00 from W43. 

The stand-off distance, $\RS$, which is a representative radius of the bow shock, is related to the bow's parameters as follows, considering the balance of momentum injection from the central body and ram pressure by the inflowing gas from outside:
\be
\RS=\sqrt{\dot{m}_{\rm w} V_{\rm w} \over 4 \pi \rho V^2},
\label{eq_bshock}
\ee
where $\dot{m}$ is the mass injection rate into the shock front, and is approximated by
\be
\dot{m}\sim M_{\rm MBS}/\tau.
\ee
Here, $\tau=d/V$ and $d\sim 10$ pc is the width of the MBS, $V_{\rm w}$ is the wind velocity from the central body, and is replaced by 
$V_{\rm w}\sim \sigma_v\sim  10 $ \kms, 
and $\rho\sim 30$ \Hcc is the ambient gas density in the outskirt of GMA taken as the logarithmic mean of the HI (8 \Hcc) and molecular (100 \Hcc) densities.
The inflow gas velocity is given by  $(V_{\rm rot}-V_{\rm p})\sin\ p \sim 30$ \kms,
{\bf 
where $V_{\rm rot}$  and $V_{\rm p}$ are the rotation velocity and pattern speed of the spiral arm, respectively, and $p$ is the pitch angle of the arm. 
}

Inserting these values into equation (\ref{eq_bshock}), we obtain $\RS\sim 54$ pc.  

\begin{figure} 
\begin{center}     
\includegraphics[width=7cm]{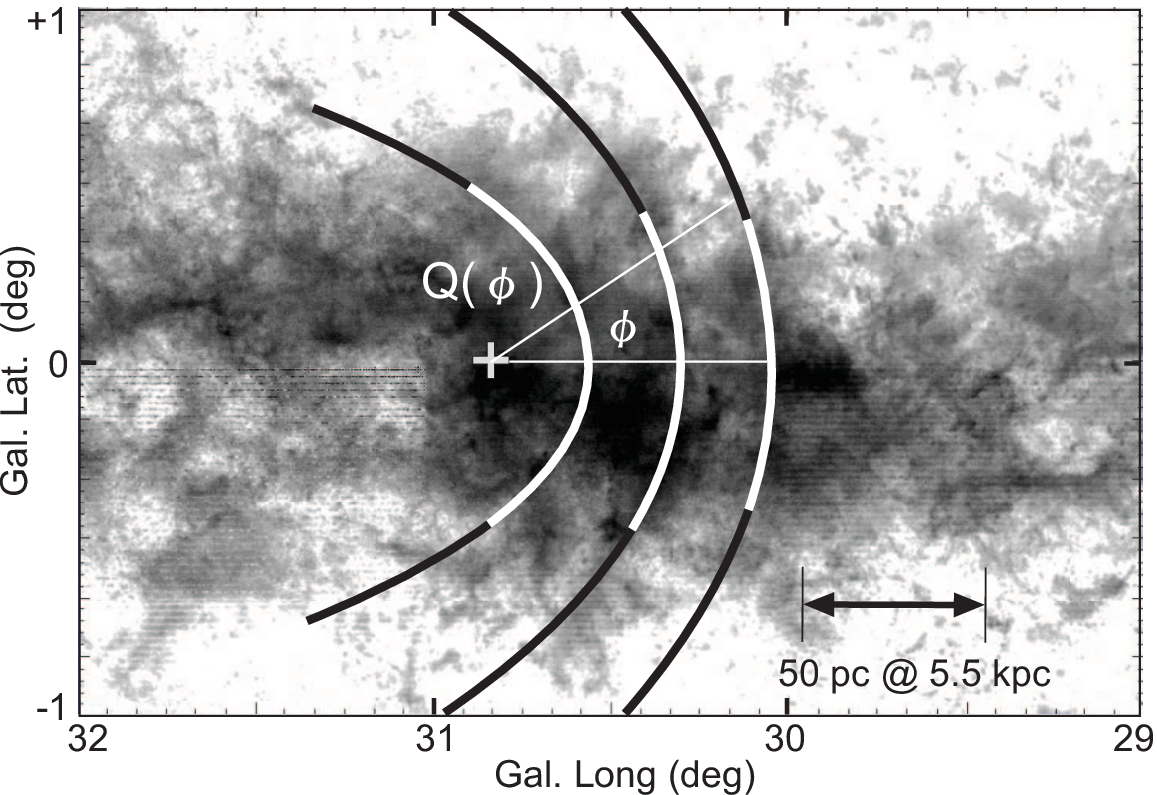}  
\end{center}
\caption{ Theoretical bow shock front for standoff distance $R_{\rm S}=25$, 50 and 75 pc centered on W43 at $(X,Y)=(0,0)$ pc, overlaid on the \co map}
\label{bshock_theory}
\end{figure}

 In figure \ref{bshock_theory} we present the bow shock shapes calculated for $\RS=25$, 50 and 75 pc overlaid on the \co intensity map, where the curve for 50 pc well represents the observed MBS shape in agreement with the above theoretical estimation of $\RS\sim 54$ pc. 
We emphasize that the bow-shock model can well represent the observed shapes and dimensions of the MBS, if the W43 and its surrounding dense molecular cloud is the triggering source of the shock. 
Full discussion of bow-shock formation mechanism in interstellar gas clouds in galactic spiral shock waves, including analyses of a number of dark-cloud bows found in nearby spiral galaxies, will be presented in a separate paper (Sofue 2018b).

\subsection{Giant commetary HII region with bright rim} 

Figure \ref{maps} shows a positional coincidence of the 10 GHz continuum bow structure with the molecular bow at G30.5, while continuum is slightly inside the MBS. The radio bow is also clearly visible on the 2,7 and 5 GHz maps using the Bonn 100-m telescope (F{\"r}st et al. 1990; Altenhoff et al. 1979) and on the 330 MHz map by Subrahmanyan \& Goss(1996) using the VLA. Spectral indices between 10 and 2.7 GHz indicated that the emission is thermal (Sofue 1985), which is also confirmed by comparing the 10 GHz intensities with those at 330 MHz considering the missing flux in the VLA map.  
 
The emission measure of the responsible ionized gas is estimated to be $EM\sim 7\times 10^3$ pc cm$^{-6}$ for assumed electron temperature of $\Te=10^4$ K (Sofue 1985). The electron density is, then, estimated to be $\ne \sim 8$ cm$^{-3}$ for an assumed line-of-sight depth equal to the vertical extent of $L \sim 100$ pc. The HII mass is calculated to be $M\sim \mH \ne L^2 W \sim 2 \times 10^4 \Msun$, which is $\sim 3\times 10^{-2}$ times the molecular mass. This yields thermal energy density, or internal pressure, of the HII gas of $p_{\rm int} \sim  \ne k \Te \sim 1.1 \times 10^{-11} $ \ergcc.  

We may consider two possible sources to ionize the gas in the continuum bow. One possible mechanism is ionization by UV illumination from inside by OB stars in W43, and the other is shock induced ionization by the inflowing supersonic flow by the galactic shock wave. 

The first, probably most likely,  mechanism is ionization from inside (down-stream side) by UV photons from OB stars in W43. This postulates the standard formation mechanism of an HII region around an OB star cluster. 
The radius of a steady state (well evolved) HII region is given by 
\be
 \rhii \simeq \left({3 \nuv \over 4 \pi \ni\ne\ar }\right)^{1/3} 
 \ee
where $\nuv$ is the UV photon number radiated by the OB stars and $\ar \sim 4\times10^{-13}  {\rm cm^{-3}s^{-1}}$ is the recombination rate.
If we assume that the luminosity of the central cluster is comparable to the far-infrared luminosity, $L\sim 1.23 \times 10^7 \Lsun$, of dust clouds in the central 10 pc of W43 (Lin et al. 2016) and that the ionized hydrogen density of $\ni\sim \ne\sim 8$ cm$^{-3}$ from the continuum EM, then, we have $\rhii \sim 130$ pc. 

Thus, W43 is luminous enough to blow off most of the ambient HII gas to radius $\sim 100$ pc. However, the expanding HII gas is blocked by the dense molecular gas blowing from the up-stream side in the galactic shock wave. This counter flow compresses the HII sphere to keep its radius smaller than $\rhii$. The observed standoff distance $\RS \sim 50$ of the MBS can be thus naturally explained by such compressed radius of HII region excited by W43. On the contraty, on the down-stream side of W43, the gas flows away from W43 causing suppressed ram pressure, so that the HII gas expands farther into the inter arm. 

\begin{figure} 
\begin{center}      
\includegraphics[width=5.5cm]{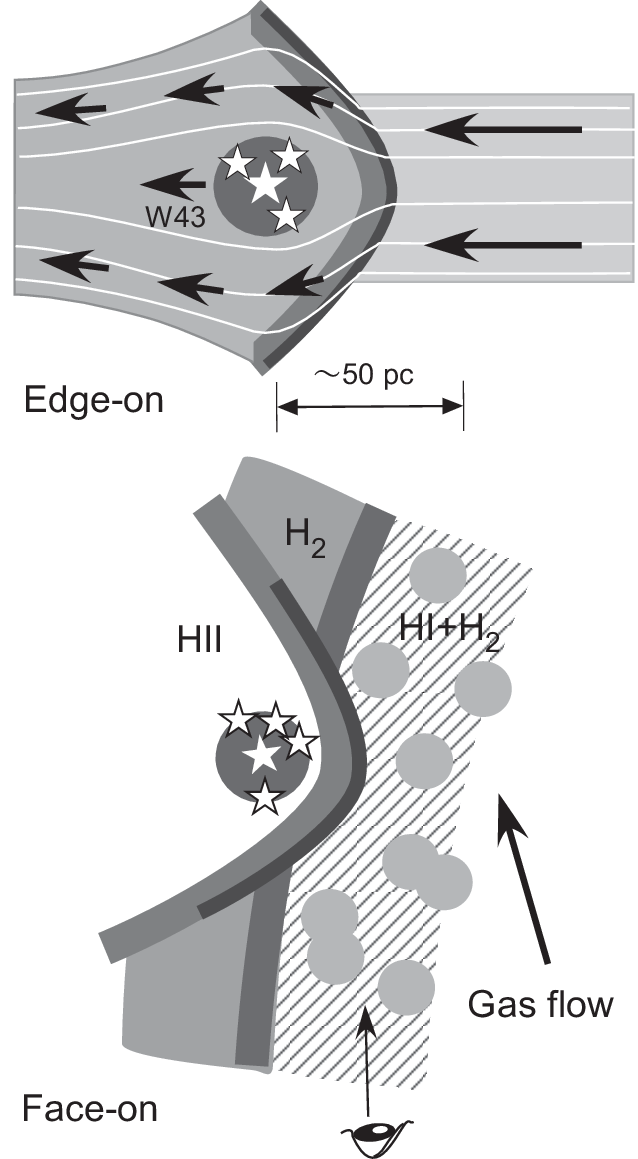}   
\end{center}
\caption{[Top] Illustration of the cross section and flow lines through the MBS. The arched shape is formed by the bow shock and hydraulic-jump in the $z$ direction. [Bottom] Same, but seen face-on. Molecular bow is formed around a giant cometary HII region, concave to the central OB stars. Similar dark bows and giant HII cones are commonly observed in star-forming spiral arms of external galaxies (Sofue 2018b).}
\label{illust_bow}
	\end{figure}

As a consequence, a giant cometary HII region showing lopesided cone of ionized gas open to down-stream side is produced inside the MBS, as illustrated in figure \ref{illust_bow}.
In fact, the radio continuum bow representing the HII rim is observed slightly inside the MBS in figure \ref{maps}. 
We also point out that similar giant cometary HII regions with H$_\alpha$-bright rim are often observed in galactic shock wave arms of external galaxies (Sofue 2018b).

Another hypothesis attributes the origin to ram pressure by the inflowing gas from the up-stream side. The ram pressure is estimated to be  $p_{\rm ext} = \mH n_{\rm pre} \Delta v^2 \sim 7\x 10^{-11}$ \ergcc  for $\Delta v \sim 10$ \kms. Here, the pre-shock gas density was estimated by HI density observed toward $\sim$G30.2+0.2 using the HI map by Bihr et al. (2015) to be $n_{\rm pre}\sim N_{\rm HI}/L\sim 40$ \Hcc, where $L\sim 100$ pc is the line-of-sight depth and $N_{\rm HI}\sim 100 \Msun\ {\rm pc}^{-2}$ is the HI column after subtraction of the extended background. Thus the inflowing gas pressure is sufficient to heat and partially ionize the gas at the bow shock. Note also that the pressure is comparable to the internal pressure by the molecular gas in the GMA of $p_{\rm mol} \sim 2 \mH n_{\rm H_2} \sigv^2 /2 \sim 7\times  10^{-11}$ \ergcc for $\sigv\sim 10$ \kms, so that it can compress the molecular gas in GMA stacked at the galactic shock in the potential well. 

The morphology of MBS and continuum bow remind us of the cometary HII regions of parsec scales, which are considered to be produced by interaction of HII gas expanding in inhomogeneous and/or flowing ISM (Reid and Ho 1985; van Buren et al. 1990; Arthur et al. 2006; Steggles et al. 2016).
We also point out that the radio bow is similar to sub-pc scale bright-rimmed clouds, where the ionization occurs due to UV photons from nearby forming stars (Urquhart et al. 2006; Sugitani et al. 1991; Thompson et al. 2004).
Although the present MBS is caused by an encounter with the supersonic flow in galactic rotation, the similarity suggests that the MBS would be a giant cometary HII region with bright-rimmed molecular gas.

\subsection{Hydraulic jump}

{\bf
Another feature to be stressed around MBS G30.5 is the sudden increase of vertical thickness of the disk from lower to higher longitudes (figure \ref{wide}, \ref{maps}). Here, we define the thickness as an averaged separation of equal intensity contours at $\Ico=150$ \Kkms in the positive and negative latitudes in figure \ref{wide}. 
Smoothing clumpy fluctuations, the up-stream disk thickness at $l\le 30\deg$ is estimated to be $\Delta b \sim \pm 0\deg.5\simeq 50$ pc. It then increases by a factor of $\sim 2$ to $\pm 1\deg\sim 100$ pc at $l=30\deg.5\deg$ in the 4-kpc arm, and returns to $\pm 0\deg.4$ (40 pc) at $l\ge 31\deg$ in the outer inter-arm region. 
}
 
The thickness jump may be attributed both to bow-shock effect by the supersonic flow as well as to a hydraulic jump from the laminar to turbulent flow at the shocks. The height of the hydraulic jump $z$ may be estimated by the change of laminar flow's kinetic energy to turbulent energy, which is on the order of 
\begin{equation}
z\sim  \eta [(V_{\rm rot}-V_{\rm p})\ \sin \ p]^2 /K_z,
\label{hydraulic}
\end{equation}
where the velocity term $\sim (V_{\rm rot}-V_{\rm p})\ \sin \ p$ is nearly equal to $\Delta V$ as observed.
Here, $\eta$ is the efficiency of conversion of laminar to turbulent kinetic energies,  and $K_z$ is the vertical acceleration by the disk's gravity. 

If we adopt $V_{\rm rot}=225$ \kms at $R\sim 4$ kpc from the rotation curve (Sofue 2013) for $V_0=238$ \kms (Honma 2015), $V_{\rm p}\sim 100$ \kms for an assumed pattern speed of $\Omega=25$ \kms kpc$^{-1}$,  $p\sim 11\deg$ (Nakanishi and Sofue 2016), and $K_z =0.76 (z/100\ {\rm pc})$ (\kms)$^{2}$ pc$^{-1}$ (Kuijken and Gilmore 1989), and assume a conversion efficiency of $\eta\sim 0.5$, then we obtain $z\sim 100 $ pc, sufficient to lift the laminar flow gas to the observed height of the MBS.

{\bf 
Variation of the disk thickness at a galactic shock wave has been investigated theoretically (Tosa 1973; Mishurov 2006) and numerically by hydrodynamic simulations (Martos and Cox 1998; G{\'o}mez and Cox 2004a, b), which showed significant increase in the disk thickness by the hydraulic jump associated with vertically extended spurs and bow-shock features.
Although the simulations qualitatively explain the observed features, the current resolution, $\sim 0.1$ kpc, is too crude to be compared with the present observation at much higher resolution, $15''$ (0.4 pc).
Quantitative comparison and more physical discussion such as to determine the efficiency $\eta$ would be a subject for the future simulations.  
}

\section{HI-to-H$_2$ transition in the Galactic Shock: Molecular Front at MBS Front}
\label{subHIH2}

Bialy et al. (2017) proposed an HI to \htwo transition scenario in the vicinity of W43 comparing the HI and \htwo column densities from 21-cm line emission and far infrared dust emission, respectively.
Their scenario may be confirmed in a more specific way by slicing the intensities into velocity channels.
 
\begin{figure} 
\begin{center}   
\includegraphics[width=7cm]{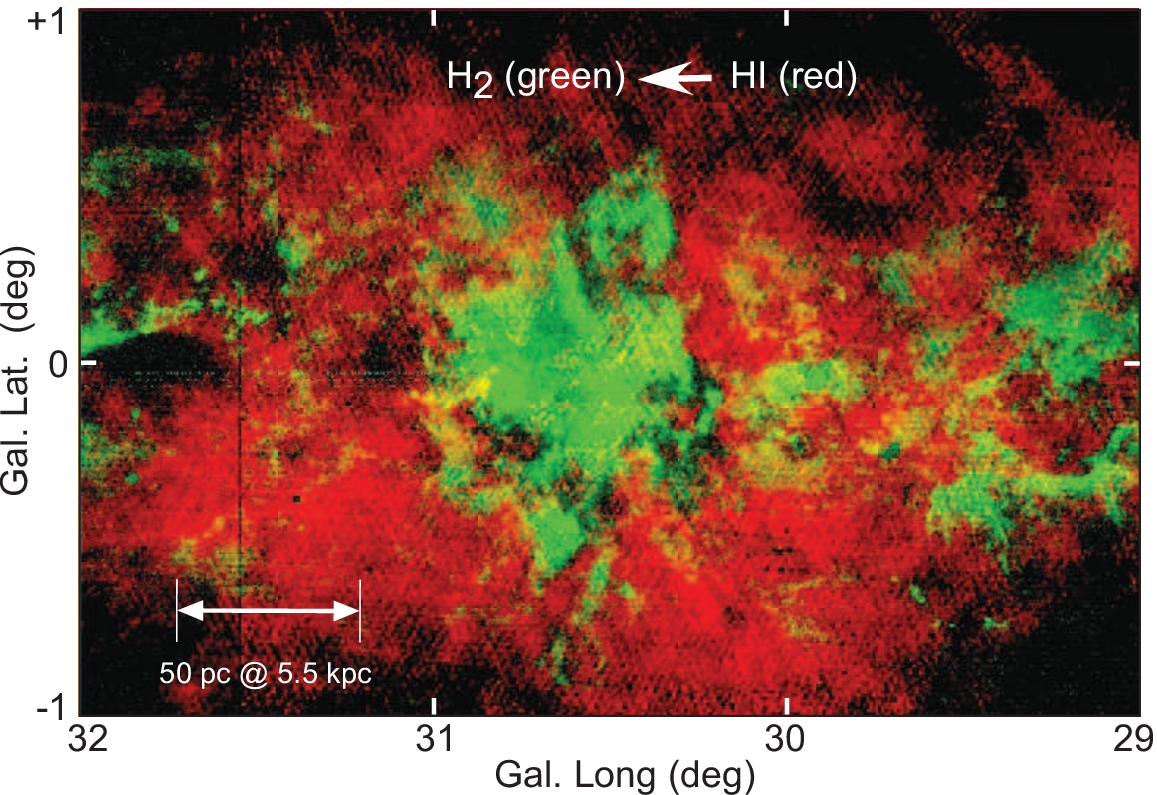}   
\includegraphics[width=7cm]{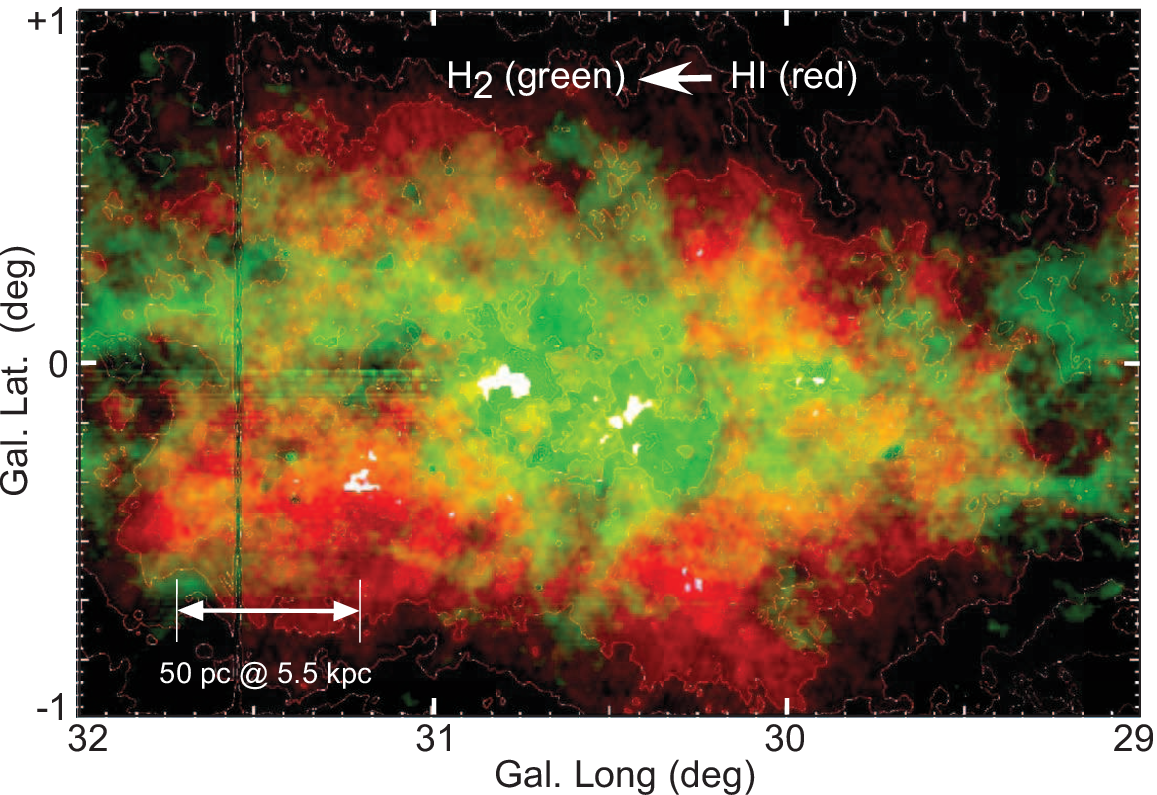}   
\end{center}
\caption{[Top] \co brightness in green overlaid on HI brightness in red at $\v=95$ \kms. [Bottom]  $\Ico$ (85 to 105 \kms) on $\IHI$ in red, made from THOR survey by Bihr et al. (2015). The intensity scales are arbitrary in order to compare the characteristic distributions in both lines. Note the deficiency of HI in CO bright regions.
}
\label{HItoH2}
	\end{figure}

In figure \ref{HItoH2} we overlay the channel maps at $\v=95$ \kms of \co brightness in green on that of HI in red, and \co intensity maps integrated from $\v=85$ to 105 \kms on the HI intensity in the same velocity range as made from the THOR HI survey by Bihr et al (2015). The figure shows that the CO strong regions are exactly located in the HI deficient regions. Note that the HI map is not corrected for absorption against continuum emission, but the absorption is significant only within $\sim 10'$ of W43 and compact HII regions.  but the effect is limited within $\sim 10'$ of W43.

We then investigate how the molecular gas is dominant in the region by mapping the molecular fraction defined by
\be
f_{\rm mol}={2\nH2 \over 2\nH2+ \nHI},
\label{eqfmol}
\ee
where $\nH2$ and $\nHI$ are the densities of \htwo and HI.
The densities are related to the observed brightness temperature through (Sofue 2018a)
\be
\nH2=\Xco \Tco {dv \/dr},
\ee
and
\be
\nHI=\XHI \Ts \tauhi {dv \/dr}.
\label{eqnhi}
\ee
where $X_i$ are the conversion factors, and $\tauhi$ is the HI optical depth defined through 
\be 
\THI= (\Ts-\Tc) (1-e^{-\tauhi}).
\label{eqTb}  
\ee
Here, $\Ts$ and $\Tc$ are the spin temperature of HI and radio continuum background temperature, respectively. Note that the velocity-related term, $d\v/dr$, disappears in equation (\ref{eqfmol}).

	\begin{figure} 
\begin{center}    
(a)\includegraphics[width=7cm]{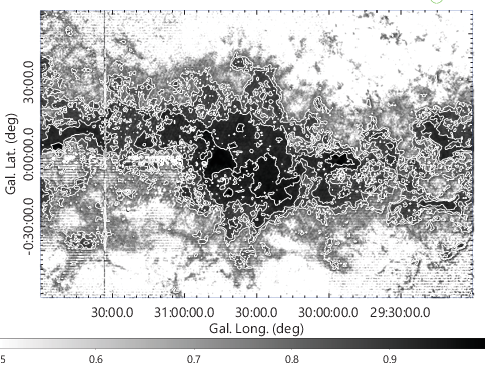}  \\ 
(c)\includegraphics[width=7cm]{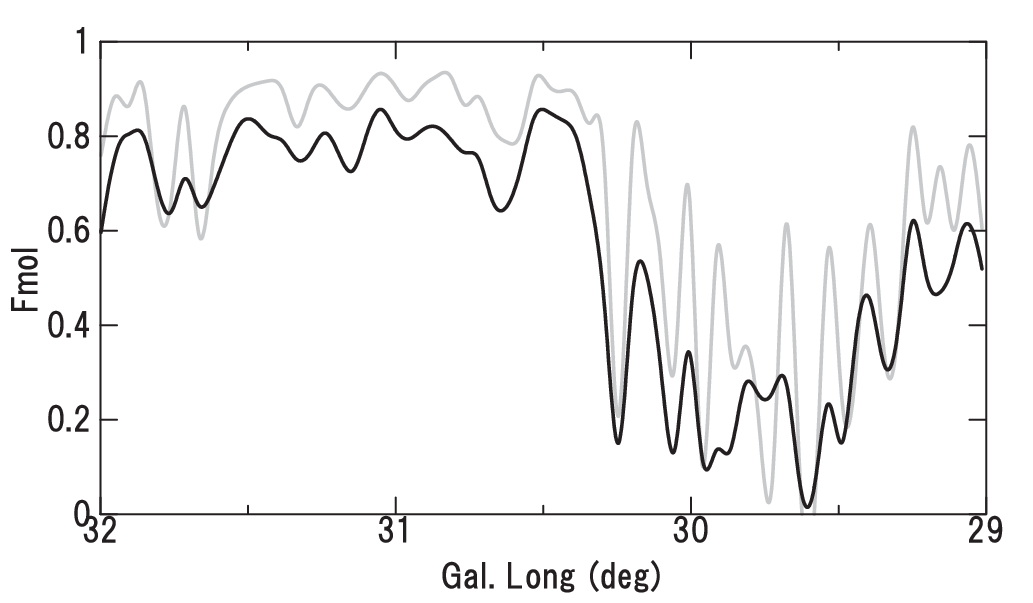}  \\
(d)\includegraphics[width=7cm]{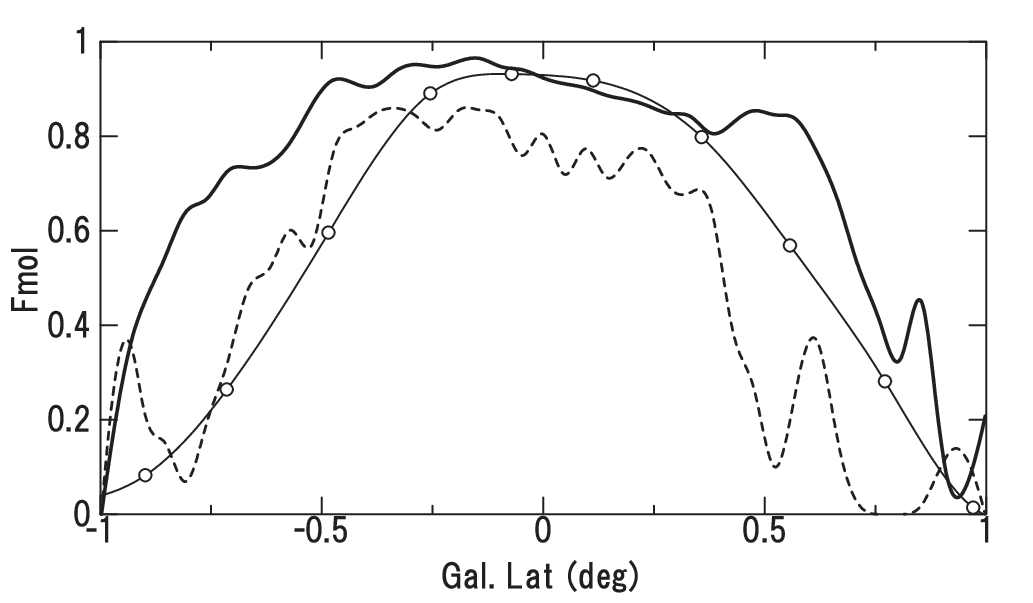}  
\end{center}
\caption{ 
(a) Molecular fraction $\fmol$. In the molecular region, the fraction is almost saturated at $f_{\rm mol}\ge 0.8$, where contours are drawn every 0.05 from 0.8 to 0.95. White regions are for below $\fmol\le 0.5$. 
(b) Horizontal variation at $b=+0\deg.5$ across MBS, showing HI to \htwo transition at the MBS by full line, and same for optically thin assumption by grey. A clear molecular front appears at the MBS front.
(c) Vertical variations at $l=29\deg.5$ (dashed line) and $30\deg.5$ (full line), showing full thickness of $0\deg.9$ (87 pc) and $1\deg.5$ (144 pc), respectively. The thin line is a global value at galacto-centric radius 4 kpc taken from Sofue and Nakanishi (2016).}
\label{fmol}
	\end{figure} 
        
Using the integrated intensity maps from 85 to 105 \kms, we obtain maps of averaged $\Tb$ maps by dividing the intensity maps by the velocity width of 20 \kms.
The averaged spin temperature at $l\sim 30\deg$ was measured to be $\Ts\sim 100$ K (Sofue 2018a). However, the observed brightness around G30.5 often exceeds this value, and we here adopt a spin temperature of $\Ts\simeq 130$ K as the maximum $\Tb$ in the map, which is considered to be saturated, and hence represent the spin temperature in the analyzed region. 
The background continuum brightness is assumed to be on the order of 1/2 of the observed brightness, and is measured to $\Tc \sim 10 K$ near the Galactic plane in the analyzed region (Sofue 2018a), small enough compared to $\Ts$. Therefore, we neglect the continuum contribution, and approximate the optical depth by
\be
\tauhi\simeq -{\rm ln} \left(1-{\THI\/\Ts} \right).
\ee  
Note that this approximation does not hold toward strong continuum sources, but their areas are negligibly small compared to the analyzed region and do not affect the result significantly. Using the thus determined optical depth, we can calculate the HI density by equation (\ref{eqnhi}).

Figure \ref{maps}(c) shows the obtained map of the optical depth. The volume density of HI is obtained by multiplying $\XHI \Ts dv/dr$ to this map, which is representatively 7.8 \Hcc for $\Ts=130$ K and $dv/dr\sim 0.1$ \kms kpc$^{-1}$. Therefore, the typical value of $\tau\sim 2- 3$ yields $\sim 12-18$ \Hcc. Note that the values toward strong continuum sources brighter than 130 K (W43, G29.96, etc.) are not valid, where the present approximation of low continuum background does not hold.

Figure \ref{fmol}(b)  shows a map of $\fmol$, and (c) and (d) are horizontal cross sections at $b=+0\deg.5$ and vertical at $l=29\deg.5$ and $30\deg.5$, respectively. The figures show that the molecular fraction is saturated in the GMA and GMC at $f_{\rm ml}\ge 0.8-0.9$. 

The horizontal cross section at $b=0\deg.5$ (figure \ref{fmol}(b)) shows sudden increase of $\fmol$ making a sharp molecular front coincident with the MBS front. This fact shows that the HI gas is transformed to \htwo by the shock compression at the MBS front.

The separate structure between CO and HI is understood by transition of HI to \htwo at the MBS edge in a short time scale.
The transverse velocity of the inflow of HI gas is on the order of $\delta V =(V_{\rm rot}-V_{\rm p}) \sin\ p \sim 30$ \kms. Then the transition time is estimated to be $t\sim \lambda/\delta V \sim 3\times 10^5$ y, where $\lambda \sim 10$ pc is the width of the transition region, which may be approximated by the width of the molecular BS.  

The constantly high molecular fraction of $\fmol\sim 0.8-0.9$ near the galactic plane manifests the global high molecular fraction in the inner Galaxy. A study of galactic-scale $\fmol$ variation indicated $\fmol\sim 0.8-0.9$ at $R\sim 4$ kpc within the molecular disk of thickness $\sim \pm 50$ pc (Sofue and Nakanishi 2017).

\section{Discussion}
\label{discussion}
\subsection{Formation mechanism of a galactic bow shock}

\begin{figure} 
\begin{center}   
\includegraphics[width=5.5cm]{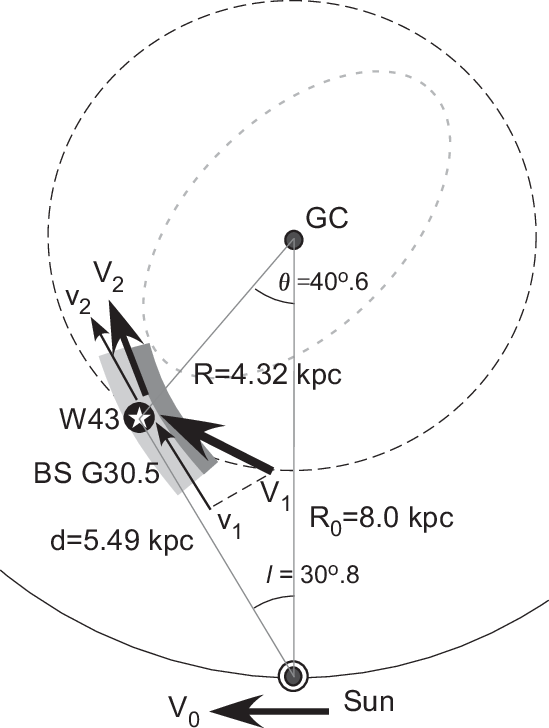} \\ 
\vskip 5mm
\includegraphics[width=7cm]{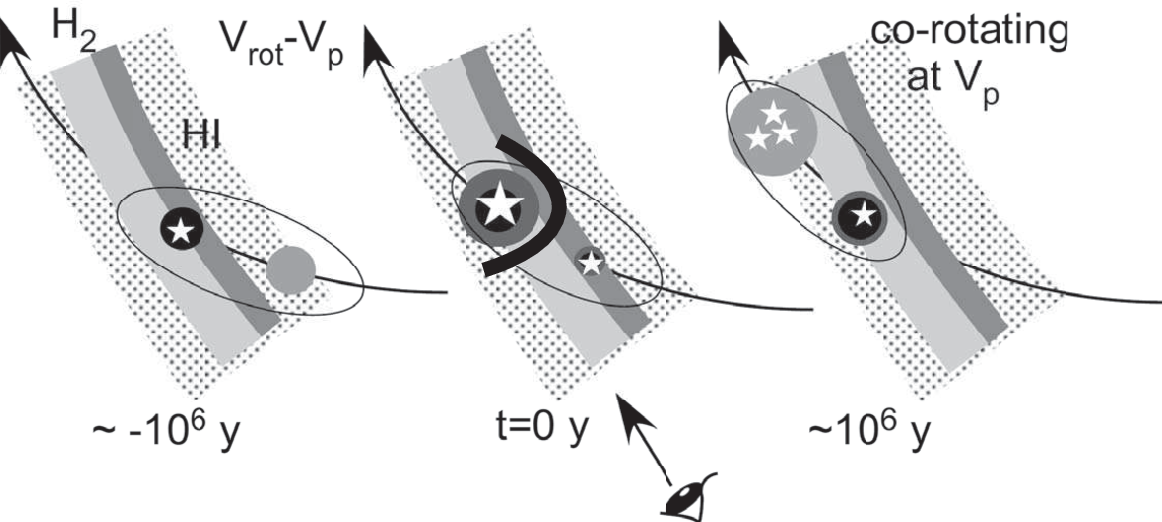}     
\end{center}
\caption{[Top] Face-on location of the 4-kpc molecular arm near W43 at a distance of 5.5 kpc adopted from Zhang et al. (2014). Thick and narrow arrows indicate the gas flow and radial velocity, respectively.  
[Bottom] Chronological passage of W43 group ($t=0$ y) in galactic rotation at $V_{\rm rot}$ through the galactic shock wave co-rotating at the pattern speed $V_{\rm p}$ of the spiral potential. The sketch is in the corotating system at $V_{\rm p}$. }
\label{illust_arm}
\label{illustEvo}
	\end{figure}

{\bf 
We first assume that the 4-kpc arm is the HI+\Htwo spiral arm defined by Nakanishi and Sofue (2016) as Arm No. 4, which is identical to the Scutum arm of Sato et al. (2015), and W43 Main, West and G30.5 MBS are located along this arm.

Since there are no parallax for W43 Main and G30.5, there remains possibility that W43 is far ($\sim 9$ kpc) and G30.5 near ($\sim 5$ kpc), or vice versa. However, the fact that MBS has a clear arc structure concave to W43 may be taken as an evidence that they are physically interactig. So, we here assume that both W43 Main and MBS are on the same side, and further, near side closer to G29.96 (West) in the same GMA. 

Based on this assumption, we consider a possible scenario to explain the kinematics and 3D molecular structure of G30.5 MBS. 
As illustrated in figure \ref{illust_arm}, the up-stream gas at velocity $V_1$ ais accelerated toward the spiral potential well, where the gas is shocked, compressed, and decelerated to velocity $V_2$. The velocity direction is bent suddenly at the front, and accordingly, the projected line-of-sight (LSR) velocity is decelerated from up- to down-stream sides.  }
        
By the bow-shock and the hydraulic-jump, the gas is also accelerated in the $z$ direction as well, and is lifted to higher $z$. According to the loss of angular momentum by the interaction with the slowly rotating spiral potential at $V_{\rm p}$, the rotation velocity, $V_{\rm rot}$, of the shocked gas is decelerated. This results in a bow-like behavior in the $v,b$ diagram (figure \ref{illust_vb}). 

The velocity gradient (figure \ref{lv_jump}) at the bow front is as large as $\sim -5$ to $-10$ \kms per $\sim 10$ pc, or $\sim -50$ to $-100$ \kms kpc$^{-1}$. This is significantly greater than the radial velocity gradient due to galactic rotation, which is on the order of $\sim -10$ \kms kpc$^{-1}$. Thus, the velocity jump can be reasonably attributed to a specific change in the local flow velocities around the BS.

According to the here proposed scenario, W43 and its molecular complex were formed prior to the encounter of the gas in the BS. This means, in turn, that the SF and molecular complex W43 are being affected by the inflowing gas from the up-stream side.

Although we assumed a spiral structure, we comment on a possible effect of the galactic bar. The noncircular motion parallel to the bar, apparently outward from the GC, would encounter the circular flow in the spiral arms (Contopoulos et al. 1989; Teuben and Sanders 1985; Kenney and Lord 1991). This may cause enhanced compression of the gas over the compression by spiral density wave, and the above scenario may work more efficiently.  
        
\subsection{Chronology in the 4-kpc arm}

Based on the time scales estimated for the clouds and for the evolutionary stages of the HII regions, we try to present a chronological view of the molecular complex centered on G30.8 embedding the HII regions and GMC. \\

\noi{\it Molecular association and clouds}

From table \ref{tabmass} we find that the detected clouds (GMA, GMC, DMC anc cores) are gravitationally bound. However, the Jeans times indicate that only larger-sized clouds than GMC can survive during the travel from one arm to the next. Smaller size and higher density clouds such as dense MC and cores are unstable and collapse within a couple of My, and are considered to be formed rather recently inside the complex during the passage of the galactic shock wave.\\

\noi{\it W43 Main (G30.8)}

The prominent HII region W43 Main is embedded in a GMC with peak CO intensity as high as $I_{\rm CO:\ peak}\sim 500$ K \kms, and the age is $\sim 1-5$ My from stellar spectroscopy (Motte et al. 2003; Bally et al. 2010). The HI line absorption agatinst the continuum begins coincident with the radial velocity of the GMC at G30.8, showing clear association of the HII region with molecular gas. Thus, W43 may be safely regarded to be the central source of the GMC and GMA.

Because the parallax of W43 Main itself is not measured (Zhang et al. 2014; Sato et al. 2015), the accurate location on the line of sight is not known. The near side distance for the recombination-line velocity $\vlsr=91.5$ \kms indicates a closer distance of $r=5.1$ kpc (Downes et al. 1980). Note that the value adopted here, 5.5 kpc, is a mean of those of G29.96, G29.86, and G31.28 and G31.58.

A supersonic flow from the up-stream side is encountering the eastern end of the W43 GMC, where the gas is shocked and widened to high latitudes, and produces the bow structure. However, the shock wave may not be directly triggering the SF activity in W43 Main. Instead, the shock at G30.5 is being formed by the encounter of the up-stream edge of the GMC surrounding W43 with the bow shock. Accordingly, a new dense MC is being formed at the root of the bow at G30.5-0.1 with peak intensity of $\sim 400$ K \kms.\\

\noi{\it W43 West (G29.96)} 

The age of G29.96 is estimated to be $\sim 0.01 - 0.1$ My (Beuther et al. 2007; Watson et al. 1997).  It is associated with a dense GMC with peak intensity $\sim 500$ K \kms. 

The line-of-sight distance is measured rather accurately to be 5.5 kpc from trigonometric parallaxes (Zhang et al. 2014), which is consistent with the the recombination-line velocity $\vlsr =99.5$ \kms (Downes et al. 1980). On the sky, G29.96 is located $\sim 80$ pc away from W43 Main and $\sim 50$ pc from the bow shock at G30.5. Particularly, the separation by 80 pc in the supersonic flow should not allow for W43 Main and West to interact physically.

Therefore, chronologically, positionally, as well as gas-dynamically, W43 West (G29.96) will not be physically related to W43 Main, except that both belong to an extended GMA. Although some physical relation has been suggested between main and West (Nguyen Luong et al. 2011), we will not argue for further connection.

\subsection{Multiple jumps and bows}

The multiple step-like behavior of the disk thickness at G29.96, G30.5, G31.4 and G31.8 (figures \ref{wide}, \ref{maps}) may be understood as a result of sequential inflow of supersonic gas from upstream side. Figure \ref{illustEvo} illustrates the chronological behavior of the molecular complex through the galactic shock wave.

The inflowing molecular complex was shocked at its down-stream side front and formed the GMC some My ago, in which W43 was born. The complex with W43 and its GMC in the center flew down to G30.8. The shock front propagated to the present place at G30.5, where is being formed a new GMC at its root. Another supersonic flow in the far up-stream at G29.9 is now forming the dense molecular cloud at G29.96 and ultra-compact HII region, which is also associated with a hydraulic jump.

According to the scenario drawn in figure \ref{illustEvo}, the two HII regions, W43 Main and G29.96, are rotating approximately on the same orbit.
They may be embedded in a GMA enclosed by contours at $\Ico \sim 100$ \Kkms in figure \ref{maps}(c), which is gravitationally bound and can be sustained for several My as shown in table \ref{tabmass}. 

As listed in table \ref{tabmass}, molecular structures less than dense MC sizes are also gravitationally bound systems, but their Jeans time is shorter than the arm crossing time, so that they cannot remain from an arm to the other. On the other hand, the whole complex has Jeans time comparable to the arm-crossing time, and can live for galactic time. Namely, the whole complex (GMA) is able to remain as a gravitationally bound system during the galactic rotation from one arm to the other. More local condensations like MCs and dense MCs, associated with SF, are formed ruing every passage through the galactic shock wave.

We point out that the thickness jumps are mostly associated with verticaly extended molecular bows, as recognized at G29.96, G30.5, G31.4 and G31.8, among which G30.5 is most prominet and discussed here in detail. The bow at G31.8 extends toward much higher latitudes, reaching the edge of the observed region at $b\sim 1\deg$ ($\pm 100$ pc). We may speculate that the molecular bow structures are commonly associated with molecular complexes formed in galactic shock waves.

\subsection{Extragalactic giant cometary HII regions and molecular bow shocks}

Giant commetray HII regions associated with inner-\Halpha rimmed molecular bows are commonly found along borders between dark lanes and stellar-bright arms in spiral galaxies (Sofue 2018b).  
The extragalactic dark bows are located on the up-stream sides of OB associations along the arms and concave to the HII regions. The dark bows have bright inside rims in \Halpha emission illuminated by the central OB stars, making giant HII cones open to down-stream of the galactic shock wave. 
The sizes are typically from $\sim 50$ to $\sim 200$ pc, depending on the luminosity of the central OB cluster as well as on strength and density of the galactic shock wave.

Although the extragalactic bows are seen from outside the galactic plane, their similarlity in shape and size is remarkable to the bow structure of MBS G30.5 seen edge-on in the Scutum arm of the Milky Way. 
We may consider that the MBS as studied here at G30.5 may be a common giant cometary phenomenon in spiral arms not only in the Galaxy but also in galaxies. 
        
\section{Summary}
 
The ideal location of the molecular complex around W43 (G30.8) near the tangent-velocity point of the Scutum arm (4-kpc arm) made it possible to investigate the 3D structure of a galactic shock wave in the densest spiral arm of the Galaxy. 

Using the FUGIN \co data cube, we measured the sizes, masses, densities, and Jeans times of the GMA, GMC, DMC and molecular cores around G30.5, and listed the result in table \ref{tabmass}. 

The molecular complex at G30.5 is associated with a molecular bow shock (MBS) of $\sim 100$ pc long and $\sim 10$ pc wide concave to W43, which extends perpendicularly to the galactic plane in positional coincidence with the radio continuum bow structure. The bow shape is well reproduced by a theoretical bow shock model with the standoff distance of $\sim 50$ pc from W43.

The side edge on the up-stream side of MBS is extremely sharp with a growth width of 0.5 pc indicative of shock front property. 
The vertical disk thickness increases at the bow shock from lower to upper longitude, indicating a hydraulic-jump.

Based on these observations, we proposed a formation scenario of the MBS, as illustrated in figures \ref{illust_bow} and \ref{illustEvo}. The molecular gas in a GMA was condensed and accelerated by the galactic spiral potential, and encountered the molecular complex around W43 and HII region, where was formed a concave bow shock.

We also showed that the HI gas has significant deficiency in the MBS and molecular regions. The molecular fraction suddenly increases and saturated at $\fmol \sim 0.9$ at the bow shock. Considering the galactic rotation and flow of gas in the analyzed region, this may be the evidence for HI-to-\htwo transition through the galactic shock wave.

\vskip 5mm
{\bf Aknowledgements}
The authors are indebted to all the staff of NRO for their continuous support during the observations. The data analysis was partly carried out on the open use data analysis computer system at the Astronomy Data Center of the National Astronomical Observatory of Japan. 


\begin{thebibliography}{}     

\bibitem[Arce \& Goodman(2002)]{2002ApJ...575..928A} Arce, H.~G., \& Goodman, A.~A.\ 2002, \apj, 575, 928 

\bibitem[Arthur \& Hoare(2006)]{2006ApJS..165..283A} Arthur, S.~J., \& Hoare, M.~G.\ 2006, \apjs, 165, 283 
\bibitem[Bally et al.(2010)]{2010A&A...518L..90B} Bally, J., Anderson, L.~D., Battersby, C., et al.\ 2010, \aap, 518, L90 
\bibitem[Baranov et al.(1971)]{1971SPhD...15..791B} Baranov, V.~B., Krasnobaev, K.~V., \& Kulikovskii, A.~G.\ 1971, Soviet Physics Doklady, 15, 791 
\bibitem[Beuther et al.(2012)]{2012A&A...538A..11B} Beuther, H., Tackenberg, J., Linz, H., et al.\ 2012, \aap, 538, A11 
\bibitem[Bihr et al.(2015)]{2015A&A...580A.112B} Bihr, S., Beuther, H., Ott, J., et al.\ 2015, \aap, 580, A112 
\bibitem[Bolatto et al.(2013)]{2013ARA&A..51..207B} Bolatto, A.~D., Wolfire, M., \& Leroy, A.~K.\ 2013, \araa, 51, 207  
\bibitem[Carlhoff et al.(2013)]{2013A&A...560A..24C} Carlhoff, P., Nguyen Luong, Q., Schilke, P., et al.\ 2013, \aap, 560, A24 
\bibitem[Contopoulos et al.(1989)]{1989ApJ...343..608C} Contopoulos, G., Gottesman, S.~T., Hunter, J.~H., Jr., \& England, M.~N.\ 1989, \apj, 343, 608 
\bibitem[Dyson(1975)]{1975Ap&SS..35..299D} Dyson, J.~E.\ 1975, \apss, 35, 299 
\bibitem[Downes et al.(1980)]{1980A&AS...40..379D} Downes, D., Wilson, T.~L., Bieging, J., \& Wink, J.\ 1980, \aaps, 40, 379  
\r Fujimoto, M. 1966, in {\it Nonstable phenomena in galaxies}, IAU Symp. 29, ed. M. Arakeljan, p. 453.
\bibitem[G{\'o}mez \& Cox(2004)]{2004ApJ...615..744G} G{\'o}mez, G.~C., \& Cox, D.~P.\ 2004a, \apj, 615, 744 
\bibitem[G{\'o}mez \& Cox(2004)]{2004ApJ...615..758G} G{\'o}mez, G.~C., \& Cox, D.~P.\ 2004b, \apj, 615, 758  
\bibitem[Herpin et al.(2012)]{2012A&A...542A..76H} Herpin, F., Chavarr{\'{\i}}a, L., van der Tak, F., et al.\ 2012, \aap, 542, A76 
\bibitem[Handa et al.(1987)]{1987PASJ...39..709H} Handa, T., Sofue, Y., Nakai, N., Hirabayashi, H., \& Inoue, M.\ 1987, \pasj, 39, 709  
\bibitem[Honma et al.(2015)]{2015PASJ...67...70H} Honma, M., Nagayama, T., \& Sakai, N.\ 2015, \pasj, 67, 70 
\bibitem[\protect\citeauthoryear{Kalberla et al.}{2010}]{2010A&A...521A..17K} Kalberla P.~M.~W., et al., 2010, A\&A, 521, A17 
\bibitem[Kenney \& Lord(1991)]{1991ApJ...381..118K} Kenney, J.~D.~P., \& Lord, S.~D.\ 1991, \apj, 381, 118 
\bibitem[Kohno et al.(2018a)]{2018PASJ..tmp....8K} Kohno, M., Torii, K., Tachihara, K., et al.\ 2018a, \pasj,  
\bibitem[Kohno, et al 2018b]{2018PASJ in prepa} Kohno, M. et al. 2018b PASJ in preparation.   
\bibitem[Kuijken \& Gilmore(1989)]{1989MNRAS.239..605K} Kuijken, K., \& Gilmore, G.\ 1989, \mnras, 239, 605 
\bibitem[Lin et al.(2016)]{2016ApJ...828...32L} Lin, Y., Liu, H.~B., Li, D., et al.\ 2016, \apj, 828, 32 
\bibitem[Louvet et al.(2014)]{2014A&A...570A..15L} Louvet, F., Motte, F., Hennebelle, P., et al.\ 2014, \aap, 570, A15  
\bibitem[Martos \& Cox(1998)]{1998ApJ...509..703M} Martos, M.~A., \& Cox, D.~P.\ 1998, \apj, 509, 703 
\bibitem[Martos et al.(1999)]{1999ApJ...526L..89M} Martos, M., Allen, C., Franco, J., \& Kurtz, S.\ 1999, \apjl, 526, L89 
\bibitem[\protect\citeauthoryear{McClure-Griffiths et al.}{2009}]{2009ApJS..181..398M} McClure-Griffiths N.~M., et al., 2009, ApJS, 181, 398 
\bibitem[\protect\citeauthoryear{Mishurov}{2006}]{2006ARep...50...12M} Mishurov Y.~N., 2006, ARep, 50, 12 
\bibitem[Motte et al.(2014)]{2014A&A...571A..32M} Motte, F., Nguy{\^e}n Luong, Q., Schneider, N., et al.\ 2014, \aap, 571, A32 
\bibitem[Motte et al.(2003)]{2003ApJ...582..277M} Motte, F., Schilke, P., \& Lis, D.~C.\ 2003, \apj, 582, 277  
\bibitem[Nguyen Luong et al.(2011)]{2011A&A...529A..41N} Nguyen Luong, Q., Motte, F., Schuller, F., et al.\ 2011, \aap, 529, A41 
\bibitem[Ogura(1995)]{1995Ap&SS.224..151O} Ogura, K.\ 1995, \apss, 224, 151 
\bibitem[Povich et al.(2008)]{2008ApJ...689..242P} Povich, M.~S., Benjamin, R.~A., Whitney, B.~A., et al.\ 2008, \apj, 689, 242 
\bibitem[Reid \& Ho(1985)]{1985ApJ...288L..17R} Reid, M.~J., \& Ho, P.~T.~P.\ 1985, \apjl, 288, L17 
\bibitem[Reipurth et al.(2002)]{2002AJ....123..362R} Reipurth, B., Heathcote, S., Morse, J., Hartigan, P., \& Bally, J.\ 2002, \aj, 123, 362 
\bibitem[Roberts(1972)]{1972ApJ...173..259R} Roberts, W.~W., Jr.\ 1972, \apj, 173, 259 
\bibitem[Sakemi et al.(2018)]{2018PASJ...70...27S} Sakemi, H., Machida, M., Akahori, T., et al.\ 2018, \pasj, 70, 27 
\bibitem[Saral et al.(2017)]{2017ApJ...839..108S} Saral, G., Hora, J.~L., Audard, M., et al.\ 2017, \apj, 839, 108 
\bibitem[Sato et al.(2014)]{2014ApJ...793...72S} Sato, M., Wu, Y.~W., Immer, K., et al.\ 2014, \apj, 793, 72 
\bibitem[Sofue(1985)]{1985PASJ...37..507S} Sofue, Y.\ 1985, \pasj, 37, 507 
\r Sofue, Y. 2013, PASJ 65, 118
\bibitem[Sofue(2018a)]{2018PASJ..tmp...57S} Sofue, Y.\ 2018a, \pasj,  in press
\r Sofue, Y. 2018b in preparation.
\bibitem[Sofue \& Nakanishi(2016)]{2016PASJ...68...63S} Sofue, Y., \& Nakanishi, H.\ 2016, \pasj, 68, 63  
\bibitem[Steggles et al.(2017)]{2017MNRAS.466.4573S} Steggles, H.~G., Hoare, M.~G., \& Pittard, J.~M.\ 2017, \mnras, 466, 4573 
 \bibitem[Subrahmanyan \& Goss(1996)]{1996MNRAS.281..239S} Subrahmanyan, R., \& Goss, W.~M.\ 1996, \mnras, 281, 239  
\bibitem[Sugitani et al.(1991)]{1991ApJS...77...59S} Sugitani, K., Fukui, Y., \& Ogura, K.\ 1991, \apjs, 77, 59 
\bibitem[Terada et al.(2012)]{2012PASJ...64..138T} Terada, Y., Tashiro, M.~S., Bamba, A., et al.\ 2012, \pasj, 64, 138 
\bibitem[Teuben \& Sanders(1985)]{1985MNRAS.212..257T} Teuben, P.~J., \& Sanders, R.~H.\ 1985, \mnras, 212, 257 
\bibitem[Thompson et al.(2004)]{2004A&A...414.1017T} Thompson, M.~A., White, G.~J., Morgan, L.~K., et al.\ 2004, \aap, 414, 1017 
\bibitem[Tosa(1973)]{1973PASJ...25..191T} Tosa, M.\ 1973, \pasj, 25, 191  
\bibitem[Ueta et al.(2008)]{2008PASJ...60S.407U} Ueta, T., Izumiura, H., Yamamura, I., et al.\ 2008, \pasj, 60, S407 
\bibitem[Umemoto et al.(2017)]{2017PASJ...69...78U} Umemoto, T., Minamidani, T., Kuno, N., et al.\ 2017, \pasj, 69, 78 
\bibitem[Wilkin(1996)]{1996ApJ...459L..31W} Wilkin, F.~P.\ 1996, \apjl, 459, L31
\bibitem[Urquhart et al.(2006)]{2006A&A...450..625U} Urquhart, J.~S., Thompson, M.~A., Morgan, L.~K., \& White, G.~J.\ 2006, \aap, 450, 625 
\bibitem[van Buren et al.(1990)]{1990ApJ...353..570V} van Buren, D., Mac Low, M.-M., Wood, D.~O.~S., \& Churchwell, E.\ 1990, \apj, 353, 570 
\bibitem[Zhang et al.(2014)]{2014ApJ...781...89Z} Zhang, B., Moscadelli, L., Sato, M., et al.\ 2014, \apj, 781, 89 
 
\end{thebibliography}
\end{document}